\documentstyle[preprint,prb,aps,epsfig,eqsecnum]{revtex}

\newcommand{\be}{\begin{equation}}
\newcommand{\ee}{\end{equation}}
\newcommand{\bn}{\begin{displaymath}}
\newcommand{\en}{\end{displaymath}}
\newcommand{\bs}{\begin{eqnarray}}
\newcommand{\es}{\end{eqnarray}}

\newcommand{\tri}{\bigtriangleup}
\newcommand{\nn}{\nonumber}
\newcommand{\Pf}{\text{Pf}\;}
\newcommand{\up}{{\mathsf up}}
\newcommand{\ri}{{\mathsf right}}
\newcommand{\dw}{{\mathsf down}}
\newcommand{\lf}{{\mathsf left}}
\newcommand{\bigstar}{\ast}
\newtheorem{theo}{Theorem}

\begin{document}

\draft
\tightenlines

\title{ Dimers and the Critical Ising Model on Lattices of genus $>1$}

\author{Ruben Costa-Santos
\footnote{e-mail rcostas@insti.physics.sunysb.edu} and Barry~M.~McCoy
\footnote{e-mail mccoy@insti.physics.sunysb.edu}}     

\address{C.N. Yang Institute for Theoretical Physics,\\
 SUNY Stony Brook, NY 11794-3840}

\preprint{YITP-SB-01-14}% [hep-th/0109167]}

\maketitle

\begin{abstract}

We study the partition function of both  Close-Packed Dimers and the 
Critical Ising Model on a square lattice embedded on a genus two surface.  
Using numerical and analytical methods we show that the determinants 
of the Kasteleyn adjacency matrices have a dependence on the boundary conditions that, for large lattice size, can be expressed in terms of genus two theta functions. 
The period matrix  characterizing the  continuum limit of the lattice is computed using a discrete holomorphic structure.
 These results relate in a direct way the lattice combinatorics with conformal field theory, providing new insight to the lattice regularization of conformal field theories on higher genus Riemann Surfaces.

\end{abstract}

%\pacs{PACS 05.50.+q 02.40.-k 64.60.Cn}

\vspace{7cm}

\pagebreak

\section{Introduction}

When  Kaufman\cite{kauf} first evaluated the finite size partition function for the Ising model on a toroidal square lattice, she found that the trace of the transfer matrix could be expressed in a very compact form as the sum of four terms.
In later solutions of the same problem, both in the combinatorial approaches\cite{pottward,kast2} or in the closely related Grassmann variable approach\cite{plech},   the sum of four terms appears as a natural way of  expressing the torus partition function.
For the Close-Pack Dimer problem, on the same lattice, Kasteleyn\cite{kast1} showed that the partition function is given also by the sum of four terms, one of them found to vanish after explicit calculation. 

Kasteleyn\cite{kast2,kast1,kast3} developed a combinatorial method of solution\cite{mccoy,russ,tesl} that can be applied to both the Ising model and the Dimer problem. In this method the four terms are the Pfaffians of four adjacency matrices corresponding to different lattice edge orientations
\be
   Z_{g=1}(T)=  \frac{1}{2}\left( -\Pf(A_1)+\Pf(A_2)+\Pf(A_3)+\Pf(A_4) \right)  \label{zz}.
\ee 

The Pfaffian of an anti-symmetric matrix is the square root of the corresponding determinant. These determinants can be  explicitly evaluated for a square lattice with  M rows and N columns.
For the Ising model at criticality and for the Dimer problem with any dimer weights,   one of the Pfaffians vanishes and the three remaining behave asymptotically\cite{ferd,ferdfish,nashoco}, in the ${\cal N}=M N \rightarrow \infty$ limit with a fixed ratio $M/N$,  as a common bulk term times a Riemann theta function\cite{mumford}  $\theta_k(0,\tau)\,$ of even characteristic. In this limit, the partition function for both models can be written as
\be
   Z_{g=1}(T_c) =  \frac{1}{2}\left[  \left|\frac{\theta_2(0|\tau)}{\eta(\tau)}\right|^d+
                                    \left|\frac{\theta_3(0|\tau)}{\eta(\tau)}\right|^d +   
                                     \left|\frac{\theta_4(0|\tau)}{\eta(\tau)}\right|^d\right] 
       \  \exp(f_{d}{\cal N})\  \left(1+ O[(\log{{\cal N}})^3/{\cal N}] \right) \label{z1}
\ee
with $d=1,2$ for Ising and Dimers respectively,  $f_d$ being the free energy of the model. The modular parameter in the theta functions given by
\be
    \tau= \left\{\begin{array}{cl} i M\cosh{2 K_h^c }/N\cosh{2 K_v^c } & \text{  for   Ising}  \\
                            i M z_h/N z_v  & \text{  for   Dimers}  \end{array} \right. \label{tau} 
\ee
with $K_v^c/\beta$ and $K_h^c/\beta $ being the vertical and horizontal coupling constants of the Ising model at the critical point and   $z_v$ and $z_h$ the vertical and horizontal Dimer weights.

Equation (\ref{z1}) provides a bridge between  lattice  combinatorics and  conformal field theory.  For the Ising model the term in square brackets is the modular invariant partition function of the  $c=\frac{1}{2}$ conformal field theory on a torus, while for Dimers it correspond to a $c=1$ conformal field theory\cite{verver2}. The theta function dependence of the determinants of the different adjacency matrices reproduce the dependence of the determinant of the Dirac operator over the different boundary conditions (or spin structures) of the  conformal field theory\cite{alvarez1,alvarez2,verver1,verver2}. 
This kind of relation between  combinatorics  and analysis is well know in the mathematics literature in the context of the Ray-Singer theorem\cite{raysin,muller} and  it has been mentioned in the context of the Ising model on the torus by Nash and O'Connor\cite{nashoco,nashoco2}.

In this paper we study the extension of (\ref{z1}) to higher genus lattices. 
 For a  lattice embedded in a genus $g$ surface, the partition function  of both Ising and Dimers  can be expressed as the sum over the Pfaffians of $4^g$ adjacency matrices\cite{kast2,russ,tesl}.  The number of different Kasteleyn orientations $4^g$  being precisely the number of different fermion boundary conditions in the corresponding genus $g$ Riemann surface. 
We  expect  the higher genus lattice determinants to be related to the functional determinants of the conformal field theory in a way similar to the toroidal case.

While expression (\ref{z1}) has been known for the torus for over thirty years,  the corresponding higher genus case has not been studied. Simple non-orientable topologies have been considered, such as the M\"obius strip\cite{brankov,luwu1} and the  Klein bottle\cite{luwu2}, and lattices with special symmetries\cite{regge2} including the genus three Klein group lattice L(2,7)\cite{reggeze}, a lattice with 168 vertices where the thermodynamic limit cannot be considered. 
Three dimensional lattices have also been studied with the Kasteleyn formalism by embedding the three dimensional lattice on a two dimensional lattice of genus of the order of the lattice size\cite{regge3}.   
While the Kasteleyn formalism is well understood for any lattice embedded on an orientable\cite{russ} or a non-orientable\cite{tesl} surface,  explicit computations of the adjacency matrix determinants are not known for higher genus lattices.

We investigate these problems by studying Dimers and the critical Ising model on the genus two lattice shown  in Fig. \ref{fig1}. We show  numerically that the corresponding sixteen Pfaffians of adjacency matrices converge to genus two theta functions and that in the large  $\cal N$ limit the partition functions of both models can be written as 
\be
    Z_{g=2}(T_c) \simeq   \left[\sum_{i=1}^{16}\left| \Theta[i](0|\Omega)\right|^d\right] \ A\   \exp(f_{d}\cal N) \label{z2}
\ee
where $d=1,2$ for Ising and Dimers respectively. The $ \Theta[i](z|\Omega)$ are the sixteen genus two  theta functions\cite{mumford}  with half-integer characteristics, defined in section \ref{section3}. The factor A includes normalization factors and non bulk terms and is not obtainable by our method that consists in the numerical evaluation of ratios of determinants of adjacency matrices for large lattice sizes. 
The term in square brackets  reproduces the classical winding part of the corresponding conformal field theory partition function\cite{verver2,g2char}. 

While on the torus, the modular parameter (\ref{tau}) can be understood   as a weighted lattice aspect ratio, on higher genus lattices the  dependence of the period matrix $\Omega$ of equation  (\ref{z2}) on the lattice properties  is more elusive.
We will study this dependence by introducing a theory of discrete differentials on the lattice and extract the period matrix  from  solutions of the finite difference Poisson equation on that lattice. The mathematical motivation for this procedure is the  idea of approximating  smooth objects defined on a surface by combinatorial objects defined on a triangulation of that surface\cite{raysin,muller,dodziuk,dodziuk2}.

The  period matrix characterizing a surface  is evaluated in two steps, the first step is the determination of the space of harmonic differentials on that surface, the second step being the decomposition of this space into the holomorphic and anti-holomorphic sub-spaces.

Harmonic differentials are determined by Hodge decomposition\cite{farkra,hodge} starting from an inner product defined on differential forms. On a lattice embedded on a surface the role of p-forms is played by the p-cochains, which are linear functionals defined on formal sums of the lattice p-elements:  vertices, edges and faces. We will call them respectively the lattice functions, lattice differentials and lattice volume forms.  Eckmann\cite{eckmann} showed that any choice of inner product on the p-cochains gives rise to a combinatorial Hodge theory and to a definition of harmonic p-cochains. Later Dodziuk\cite{dodziuk} and Dodziuk and Patodi\cite{dodziuk2} proved that for a suitable choice of  this inner product on p-cochains, of a triangulation of a surface,  the  resulting combinatorial Hodge theory  converges, in the continuum limit, to the surface Hodge theory.

While our definitions are based on the same general idea, replacing p-forms on a surface by p-cochains on a lattice embedded on that surface, we depart from  Dodziuk and Patodi work on several points. The main difference is that their definition of the inner product on p-cochains depends on a mapping of these cochains into forms on the embedding surface\cite{whitney} while our definition is done in term of lattice quantities.  Dodziuk and Patodi  discrete  theory  is defined referring to the surface properties while we are interested in the inverse procedure: to built the continuum theory from the combinatorics of the discrete lattice.

To determine the period matrix we need the additional concept of discrete holomorphic differentials. These problems  have been considered in the study of finite element electrodynamics\cite{books,kotiuga,bourdlala,tarketboss,teix1} where it is well known that the concept of holomorphic differentials ``defy a straightforward discretization''\cite{hiptmair}.
In the context of the Ising model, discrete holomorphy  can be traced back to the finite difference equations on the correlation functions found by McCoy, Perk and Wu\cite{mccoy1} and has been  recently discussed  by Mercat\cite{mercatt,mercat}.  In Mercat's work the difficulty of defining discrete holomorphy for a finite size lattice is patent from the fact that the various spaces of differentials have dimensions double of the continuum analogues. 

In this paper we give a construction procedure for holomorphic differentials, based exclusively on the direct lattice, which becomes meaningful for large lattice size. To the numerical precision we were able to test it, this procedure has a well defined continuum limit and reproduces the period matrices observed in equation (\ref{z2}).  
This is the first time that an argument is given to evaluate modular parameters for higher genus lattices.

The structure of the paper is as follows:

In section \ref{section2}  we review the Kasteleyn formalism and specify the adjacency matrices for both the Ising model and the Dimer problem. In section \ref{section3} we establish (\ref{z2}) by showing numerically that ratios of determinants of the adjacency matrices converge to ratios of theta functions. In section \ref{section4}  we define  discrete harmonic differentials on the lattice and give a procedure for numerical evaluation of the corresponding basis. In section \ref{section5} we discuss discrete holomorphic differentials and the computation of approximations to the period matrix.   In section \ref{section6} these evaluations of lattice period matrices are compared with the period matrices of (\ref{z2}), obtained by direct fit of determinant ratios.
Our conclusions are presented in section \ref{section7} and the technical details of the Kasteleyn formalism are discussed in appendix \ref{append1}.

\section{Determinants of adjacency matrices} \label{section2}

Consider the square lattice with the boundary conditions shown in Fig. \ref{fig1}. This lattice, $G$, is characterized by five integers sizes $(M_1,M_2,K,N_1,N_2)$ and two coupling constants or dimer weights ($z_h$,$z_v$) distinguishing the vertical and the horizontal directions.
The boundary conditions are such that the lattice can be drawn without superposition of edges only on a surface of genus two or higher. The cycles of lattice edges $ a_i, b_i$, with $i=1,2$, represent a canonical basis of the first homology group of the embedding surface. The same basis drawn over the dual lattice edges will be denoted by $\tilde a_i, \tilde b_i$ (see Fig. \ref{fig2}).

We start by considering the Close-Packed Dimer problem on this lattice and for simplicity we assume that all integer sizes $(M_1,M_2,K,N_1,N_2)$ are even integers. A  close-packed dimer configuration in $G$ is a selection of edges such that every vertex is included once and only once as a boundary of an edge, see Fig. \ref{fig9} for examples. If a weight $z_h$ is assigned to the 
horizontal edges and a  weight $z_v$ to the vertical edges, the dimer partition function is defined to be
\be
    Z^{\text{Dimer}}= \sum_{\text{dimer config.}}^{G} z_h^{n_h} z_v^{n_v} 
\ee 
where $n_h$ and $n_v$ are the number of horizontal and vertical edges in a given dimer configuration and the sum runs over all possible dimer configurations on $G$.

While the Close-Packed Dimers problem is an interesting statistical mechanics model on its own  it also provides a combinatorial approach to the Ising model. There is a  well known correspondence between polygon configurations of the Ising model high temperature expansion and  dimer configurations on a decorated lattice\cite{kast2}. The Ising model partition function on the lattice $G$ can then be expressed as
\be   
  Z^{\text{Ising}} = (2 \cosh{K_v}\cosh{K_h})^{{{\cal N}}} \;\sum_{\text{dimer config.}}^{G'}  {w_v}^{n_v} {w_h}^{n_h} 
\ee
with $w_i=\tanh{\beta K_i }$,  where  $K_v$ and $K_h$ are the vertical and horizontal coupling constants and $ {\cal N}$ is the number of vertices in the lattice $G$. The sum runs over all the dimer configurations of the decorated lattice $G'$ represented in Fig. \ref{fig3}.

Kasteleyn developed a combinatorial formalism\cite{kast2,kast1,kast3,mccoy,russ,tesl} that allows the expression of these dimer partition functions as a sum over the Pfaffians of $4^g$ adjacency matrices, where $g$ is the genus of the simplest surface where that lattice can be drawn without superposition of edges.  A detailed discussion of the higher genus Kasteleyn formalism is given in appendix \ref{append1}, here we will state only the main results.

The Kasteleyn adjacency matrices are defined in the following way:  label the lattice vertices with an integer from 1 to ${\cal N}$ and choose a lattice edge orientation by assigning to each edge a direction represented by an arrow (see Figs.  \ref{fig2} and  \ref{fig4} for two examples). The signed adjacency matrix corresponding to this edge orientation is the ${\cal N}\times {\cal N}$ matrix $A_{ij}$ with entries

\be
       A_{ij}= \left\{
              \begin{array}{rl}
              z &  \textstyle{\text{ if there is an arrow from vertex i to vertex j of weight }} z \\ 
             -z &  \textstyle{\text{ if there is an arrow from vertex j to vertex i of weight }} z\nn \\ 
              0 & \textstyle{\text{ otherwise}} 
               \end{array}   \right. 
\ee

The Pfaffian of such an ${\cal N}\times {\cal N}$ anti-symmetric matrix, with ${\cal N}$ even, is defined as
\be
    \Pf  (A) = \frac{1}{2^{{\cal N}/2}({\cal N}/2)!} \sum_p \epsilon_p \;A_{p_{1}p_{2}}A_{p_{3}p_{4}}\cdots A_{p_{{\cal N}-1}p_{{\cal N}}} \label{pff} 
\ee 
where the sum goes over all the permutations $p$ of the integers from 1 to ${\cal N}$ and $\epsilon_p=\pm 1$ for even and odd permutations respectively.

{}From the definition of the adjacency matrix it is clear that each non zero term in the Pfaffian expansion equals, in absolute value, a term in the corresponding dimer partition function. The relative sign between different terms depends on the choice of the edge orientation. We would like to choose an orientation  such  that all the terms in the Pfaffian have the same relative sign. Then the Pfaffian of the adjacency matrix would equal the partition function modulo an overall sign.

Kasteleyn showed in a beautiful tour de force of combinatorics\cite{kast1} that edge orientations with this property exist. These are the edge orientations such that every lattice face was an odd number of clockwise oriented edges. In a genus $g$ lattice there will be  $4^g$ different such orientations and the dimer partition function is given as a linear combination of their Pfaffians\cite{kast3,russ,tesl}.

For the genus two lattices $G$ and $G'$, that we are considering in this paper, the relevant sixteen edge orientations can be labeled as $A(n_{\tilde a_1},n_{\tilde b_1},n_{\tilde a_2},n_{\tilde b_2})$ with $n_x=0,1$ for $x= \tilde a_1,\tilde a_2,\tilde b_1,\tilde b_2$.  The starting orientations $A(0000)$ are shown in Fig. \ref{fig2} and Fig. \ref{fig4} for the lattices $G$ and $G'$ respectively.  An orientation with a certain $n_x=1$ is obtained from the corresponding orientation with $n_x=0$ by introducing a disorder loop along the non-trivial cycle $x$, this is, by reversing the orientation of all the edges crossed by the cycle $x$. 

To  make connection with the theta functions characteristics and allow for more compact equations we will also use the alternative notations
\be
   A(n_{\tilde a_1},n_{\tilde b_1},n_{\tilde a_2},n_{\tilde b_2})= A\scriptsize{\left[\begin{array}{cc}   n_{\tilde b_1} & n_{\tilde b_2} \\ n_{\tilde a_1} & n_{\tilde a_2} \end{array} \right]}=A_i \label{orient} \label{defori}
\ee 
 with the integer label given by $i={\textstyle 16-8n_{\tilde a_1}- 4n_{\tilde b_1}-2n_{\tilde a_2}-n_{\tilde b_2}}$.

We show in appendix \ref{append1} that, in terms of these orientations, the Dimer and the Ising model partition functions on $G$ are given by
\bs
   \begin{array}{rccccccccccccccccl}
Z=\frac{1}{4}(&P_1&-&P_2&-&P_3&-&P_4&-&P_5&+&P_6&+&P_7&+&P_8&-&  \\  
     &P_{9}&+&P_{10}&+&P_{11}&+&P_{12}&-&P_{13}&-&P_{14}&+&P_{15}&+&P_{16}&)& \end{array} \label{zzz}
\es 
where $P_i=\Pf A_i$ and the adjacency matrices are defined on the lattice $G$ for Dimers and on the lattice $G'$ for the Ising model. 

For translational invariant lattices, as the torus square lattice, the Pfaffians of the adjacency matrices can be evaluated in a closed form and the theta function dependence can be extracted by a careful asymptotic analysis\cite{ferd,ferdfish,nashoco}. 
For the genus two lattices $G$ and $G'$ such an analytic treatment is not possible and we are forced to resort to numerical evaluations of the Pfaffians. 

\section{Ratios of determinants and theta functions}\label{section3}

Pfaffians of adjacency matrices can be numerically evaluated for large lattice sizes using the fact that $\Pf A\!\!=\!\!\sqrt{\det A}$ for an anti-symmetric matrix A.
For a given  lattice aspect ratio  $(m_1,m_2,k,n_1,n_2)$ we evaluate the ratios of determinants of the adjacency matrices for a sequence of lattices characterized by the integers sizes $(M_1,M_2,K,N_1,N_2)= (m_1,m_2,k,n_1,n_2)\,L$ with increasing $L$. 
The Ising model coupling constants were chosen to satisfy the criticality condition for a square lattice on the torus
\be
    \sinh(2 \beta K_h)\sinh(2 \beta K_v)=1 \label{critic}
\ee
while no restrictions were imposed on the vertical and horizontal weights of the Dimer model. Numerical examples  for the critical Ising model and Dimers are shown in tables \ref{table4} and \ref{table5}. 

The fifteen ratios  of determinants, obtained is this way, are found to converge smoothly with the lattice size $\cal N$, see Fig.  \ref{fig69}, \ref{fig70} and \ref{fig68}. The solid line in  Fig. \ref{fig69} and \ref{fig70} is a fit with a quadratic polynomial on $1/ \cal N $ to the values obtained for different lattice sizes.

Unlike the toroidal case, none of the sixteen determinants vanishes at finite size for the  (\ref{critic}) choice of the Ising coupling constants  or for any choice of vertical and horizontal dimer weights of the Dimer problem. There are choices of Ising coupling constants with small deviations from the criticality condition (\ref{critic}) that  will make some of the determinants vanish at finite size but not all of them simultaneously.
However six of the determinant ratios  converge to  small values for large $\cal N$ and can be associated with theta functions of odd characteristic. The convergence of these six ratios is shown in  Fig. \ref{fig68} where the values corresponding to the Ising model ratios are squared for comparison.

The genus two theta functions $\Theta$  are defined by the quickly converging series expansion \cite{mumford} 
\be
   \Theta {\left[\begin{array}{c}  { \mbox{\boldmath $\alpha$}} \\  { \mbox{\boldmath $\beta$}} \end{array} \right]}  \left({\bf z},\Omega \right)
   =\sum_{{\bf n}\in Z^2} \exp{\left[i\pi({\bf n}+ { \mbox{\boldmath $\alpha$}})^T \Omega({\bf n}+{ \mbox{\boldmath $\alpha$}})+2\pi i({\bf n}+{ \mbox{\boldmath $\alpha$}})^T ({\bf z}+{ \mbox{\boldmath $\beta$}})\right]}   \label{theta}
\ee
where {\boldmath $\alpha$}, {\boldmath  $\beta$}, ${\bf  z}$ and ${\bf n}$ are  2-vectors half-integers, complex numbers and integers respectively. The $2\times 2$ period matrix $\Omega$ is a symmetric complex matrix with positive definite imaginary part.

We found that the extrapolated values of the determinants ratios, in the  $L\rightarrow \infty$  limit, can be expressed in terms of ratios of theta functions as
\be
  \left. \frac{\det\left( A\scriptsize{\left[\begin{array}{cc} c_1&c_2\\c_3&c_4\end{array} \right]}\right)}
             {\det \left(A\scriptsize{\left[\begin{array}{cc}0 & 0\\ 0 & 0 \end{array} \right]}\right)} \,\right|_{T_c} =
  \left(\frac{\Theta\scriptsize{\left[\begin{array}{cc} c_1/2&c_2/2\\c_3/2&c_4/2  \end{array} \right]}\left(0,\Omega \right)}
             {\Theta\scriptsize{\left[\begin{array}{cc}  0&0\\ 0&0 \end{array} \right]}\left(0,\Omega \right)}\right)^d \label{resultati}
\ee
for the 16 combinations of $c_i=0,1$  with $d=2$ for critical Ising and $d=4$ for Dimers. The period matrix $\Omega$ being determined from the determinant ratios by a suitable numerical fitting procedure. See  tables \ref{table8} and  \ref{table9} for examples. Each table displaying the results for three different lattices, the first column shows  the  $L\rightarrow \infty$ extrapolated ratios of determinants and the second column shows the theta ratios
\be
     \Theta_{(16-8d_1-4c_1-2d_2-c_2)}(\Omega)=\Theta{\scriptsize \left[\begin{array}{cc} c_1/2& c_2/2\\ d_1/2& d_2/2\end{array} \right]}\left(0,\Omega \right)/\Theta{\scriptsize\left[\begin{array}{cc} 0&0\\ 0&0\end{array} \right]}\left(0,\Omega \right) 
\ee
for $c_i,d_i=0,1$,  with a period matrix obtained by numerical fit. The precision to which the two sets of numbers agree is remarkable, typically a precision from $10^{-4}$ to $10^{-6}$.

For the Ising model, way from the criticality condition (\ref{critic}), the sixteen determinants are found to converge rapidly to the same value, while for the Dimer problem the theta function expression (\ref{resultati}) is valid for all values of the dimer weights.

The period matrices found in (\ref{resultati})  are  purely imaginary for all lattice aspect ratios and coupling constants. For both Ising and Dimers we have
\be
     \Omega=i\left[\begin{array}{cc} \Omega_{11} &\Omega_{12}\\\Omega_{12}& \Omega_{22}\end{array} \right].
\ee
For such a period matrix the theta functions at zero argument are real. This seems to be a property of locally square lattices observed on the toroidal square lattice (\ref{tau}) but not on the corresponding triangular lattice\cite{nashoco} where the modular parameter is in general complex.

\section{Harmonic differentials on the lattice}\label{section4}

In this section we will consider quantities defined on the lattice p-elements: vertices, edges and faces.  We will call them respectively the lattice functions,  lattice differentials and  lattice volume forms.

A lattice function $f$   is defined by its value at each lattice 
vertex and can be represented by a ${\cal N}$-vector $f[n]: n= 1,\ldots,{\cal N}$ after 
some integer labeling of the ${\cal N}$ vertices of the lattice is chosen.

A lattice differential $w$ is defined by its value on the oriented lattice edges and will be represented by a ${\cal N}\times 2$-matrix $w[n|p]: n=1,\ldots,{\cal N}; \ p=1,2$ where $[n|1]$ stands for the edge immediately right of the vertex $n$ and $[n|2]$ for the edge immediately above, referring to the lattice drawing of Fig. \ref{fig1}. We define the horizontal edges to be oriented from left to right and vertical edges from bottom to top. The line integral of a lattice differential $w$ along a path  $C$ of lattice edges is the sum of the values of $w$ on all the edges included in the path 
\be
   \int_{C} w= \sum_{[n|p]\in C } \pm \,w[n|p]
\ee
the minus sign applying  to edges with opposite orientation to the one of the path. 

A lattice volume form $\eta$ is defined by it´s value on each lattice face. The lattice in Fig. \ref{fig1} when drawn on a genus two  surface has $ {\cal N}-2$ faces,  two of which are octagons the remaining being squares.  The volume form $\eta$ will be
represented by a (${\cal N}$-2)-vector $\eta [q]: q= 1,\ldots,{\cal N}-2$ where $q$ is an integer labeling the ${\cal N}$-2 faces of the lattice. The integral of a volume form $\eta$ over a given area $A$ on the lattice is the sum of the values that $\eta$ takes on all the lattice faces contained in that area
\be
   \int\!\!\!\int_{A} \eta= \sum_{[q]\in A }  \eta[q]
\ee

It is convenient to relate the labeling of vertices with the labeling of faces. For that purpose we introduce the following notation: if $q$ labels a squared  face then $\hat q_1$ stands for the label of its lower left vertex; if $q$ labels an octagonal face then $\hat q_1$ and $\hat q_2$ stand for the labels of the two vertices that can be seen as the octagons lower left vertices in Fig. \ref{fig1}. Reciprocally if $n$ labels a lattice vertex then $\tilde n$ is the label of the lattice face of which $n$ can be seen as the lower left vertex. We then have the relations
\bs
   \tilde{\hat{q}_i}&=&q \\
    n &\in & \{ \hat{\tilde n}_1 ,\hat{\tilde n}_2  \}
\es

An inner product on lattice functions,  lattice differentials and  lattice volume forms can be defined in the following way
\bs
    (f,f') &=&  \sum_n  f[n]f'[n] \\ \label{cinn}
    (w,w') &=&  \sum_n ( \ (h/v)\;w[n|1]\,w'[n|1]+ (h/v)^{-1}\; w[n|2]\,w'[n|2]) \nn \\
    (\eta,\eta ') &=&  \sum_q  \eta[q]\eta '[q] \nn   
\es
where $h/v$ is a positive parameter providing different weighting of the horizontal and vertical components of a differential.

The lattice exterior derivative $d$ is a linear operator that takes  lattice functions into
lattice differentials and lattice differentials into lattice volume forms. For our labeling and choice of orientations $d$ is given by
\bs
  (d\, f)\, [n|1]&=& f[\ri(n)] - f[n] \label{der}\\
  (d\, f)\, [n|2]&=& f[\up(n)] - f[n] \nn\\
 (d\, w)\, [q]&=& \sum_i \left( w[\hat q_i|1] + w[\ri(\hat{q}_i)|2] - w[\up(\hat{q}_i)|1] - w[\hat{q}_i|2]\right) \nn
\es
with $i$=1,2 for octagonal faces and $i$=1 for squared faces. The functions $\ri(n), \lf(n), \up(n)$ and $\dw(n)$  give the label of the vertex immediately right, left, above and below of vertex $n$, referring to the lattice drawing of Fig. \ref{fig1}.

The exterior derivative defined in this way satisfies a discrete version of Stokes theorem. Let $C(n,n')$ be a path of lattice edges from vertex $n$ to vertex $n'$ and $A$ an area on the lattice, then we have that
\bs
   \int_{C(n,n')} d\, f&=& f[n']-f[n]  \\
   \int\!\!\!\int_{A} d\, w&=&  \int_{\partial A} w \nn
\es 
where $\partial A$ is the path along the boundary of the area $A$ with an anticlockwise orientation.

The co-derivative $\delta$ is defined as the  adjoint operator of the exterior derivative
\bs
   (w,d\,f) &=& (\delta\, w,f) \\
   (\eta,d\,w) &=& (\delta\, \eta,w)\nn
\es
and can be expressed as an operator that takes  lattice differentials into  lattice functions and  lattice volume forms into  lattice differentials
\bs
   (\delta\,w)\,[n]&=& \frac{h}{v}\left( w[\lf(n)|1]-w[n|1]\right) + \frac{v}{h}\left( w[\dw(n)|2]-w[n|2]\right)\label{del} \\
  (\delta\,\eta\,)\,[n|1]&=& \frac{ v}{h}\left( \eta[\tilde n]-\eta[\widetilde{\dw}(n)]\right) \nn \\
  (\delta\,\eta)\,[n|2]&=& \frac{h}{v}\left( \eta[\widetilde{\lf}(n)]-\eta[\tilde n]\right) \nn \\
\es
where $\widetilde{\lf}(n)$ stands for the face given by the tilde of the vertex ${\lf}(n)$ and similar for $\widetilde{\dw}(n)$.

Both the exterior derivative and the co-derivative satisfy $dd=\delta\delta=0$. Explicitly in terms of the formulas above we have that
\bs
 (dd\, f)\, [q]&=& \sum_i \left( f[\up(\ri(\hat q_i))]-f[\ri(\up(\hat q_i))] \right)=0 ,\\
 (\delta\delta\,\eta)\,[n]&=& \eta [\widetilde{\lf}(\dw(n))]- \eta [\widetilde{\dw}(\lf(n))]=0 .
\es
There are four vertices on the octagonal faces  where ${\ri}(\up(n))\neq {\up}(\ri(n))$, at these faces  $dd$  vanishes because of cancellation between the $\hat q_1$ and the $\hat q_2$ terms.  There are also four vertices where ${\lf}(\dw(n))\neq{\dw}(\lf(n))$ but $\widetilde{\lf}(\dw(n))=\widetilde{\dw}(\lf(n))$ is true for all vertices.

In exact analogy with the continuum definitions, a  lattice differential $w$ is said to be closed if $d\,w=0$, exact if $w=d\,f$, co-closed if $\delta\,w=0$ and co-exact if $w=\delta\,\eta$. 
We are mainly interested in the  harmonic differentials, a lattice differential is said to be  harmonic if it is both closed and co-closed
\be
     w \text{   is harmonic } \equiv \left\{
              \begin{array}{cl}
                    d\, w[q] =0 ,   & \ q=1,\ldots,{\cal N}-2\\ 
                \delta\, w[n]=0 ,   & \ n=1,\ldots,{\cal N} 
              \end{array}   \right.  \label{harm}
\ee

In general, on a genus $g$ lattice, the lattice harmonic differentials form a vector space of dimension $2g$. For the genus two lattice $G$ this can be seen directly from  the system of equations (\ref{harm}). It is a system of $2{\cal N}-2$  equations on $2{\cal N}$ unknowns and there are at least two independent solutions. Two additional solutions are provided by constant differentials with independent vertical and horizontal components and the  system has  at least four independent solutions. 

It is remarkable that, on the lattice, the dimension of the space of harmonic differentials is a linear algebra problem determined by the Euler characteristic. Consider a genus $g$ lattice with the property that its edges can be distinguished into two classes: vertical and horizontal. For such a lattice the dimension of the space of harmonic differentials, that is, the number of independent solutions of the system (\ref{harm}) on that lattice, is at least
\be
    \text{\#(edges)}-\text{\#(vertices)}-\text{\#(faces)}+2= -\chi+2 =2g 
\ee
where $\chi$ is the Euler characteristic of the lattice.

It still remains to be proven that there are not more than $2g$ linear independent harmonic lattice differentials.  Note that $2g$  is the  dimension of the lattice first homology group. We will prove, in a way similar to the continuum, that a harmonic lattice differential is completely determined by its integrals, or periods, around the lattice non-trivial loops and therefore there can only be  $2g$ independent harmonic differentials.

Harmonic differentials are by definition closed and its integrals depend only on the homology class of the path considered. If two differentials $w$ and $w'$ have exactly the same  periods along the $2g$ classes of loops then their difference $w-w'$ has zero integral along any closed path on the lattice and the function
\be
   f[n]= \int_o^n\, (w-w') \label{funk}
\ee  
is well defined for any lattice path going from a fixed vertex $o$ to the vertex $n$. This function is a harmonic function, in the sense that it satisfies
\be
   \Delta f[n]\equiv(\delta d\,f)[n]=\delta\,(w-w')[n]=0\label{deflap}
\ee
with the Laplacian operator given explicitly by
\bs
   (\Delta\;f)[n]&=& 2(h/v+v/h)\,f[n]\label{lap}\\
                 && -h/v\,(f[\ri(n)]+f[\lf(n)]) \nn \\
                 && -v/h\,(f[\up(n)]+f[\dw(n)]) \nn
\es
It is easily seen that a function harmonic at $n$ cannot have a local extremum at $n$ and that the only functions  harmonic everywhere on the lattice are constants. From this it follows that the function defined in (\ref{funk}) vanishes at all lattice vertices. In this way, we proved that if two harmonic differentials $w$ and $w'$ have the same periods then they are equal, $w=w'$. The space of harmonic differentials has therefore the same dimension as the first homology group, $2g$.

For a given lattice a basis of the space of  harmonic differentials can be evaluated  by  direct solution of the linear system of equations (\ref{harm}). Such a solution can always be found, at least numerically, although the evaluation of the kernel of a large linear system is computationally demanding. For practical purposes there is a much more efficient method based on a discrete version of the Hodge decomposition theorem.

The discrete Hodge decomposition theorem states that the  space of lattice differentials  has an orthogonal decomposition in terms of exact, co-exact and harmonic differentials and that any  lattice differential $w$ can be written in an unique way as
\be
 w= d f+ h+\delta \eta \label{hdg}
\ee
where $h$ is a harmonic differential. The orthogonality of the different kinds of differentials follows directly from (\ref{harm}) and the property $dd=\delta\delta=0$.

This result allows the determination of harmonic differentials by orthogonal projections. Our objective is to obtain a basis $\{A_k, B_k :\;  k=1,2\}$ of the space of harmonic differentials satisfying the normalization conditions
\bs
    \int_{a_j} A_k&=& \delta_{kj},\ \ \int_{b_j} A_k= 0 \label{harmbasis}\\
    \int_{a_j} B_k&=& 0,\ \ \; \int_{b_j} B_k=  \delta_{kj} \nn
\es
with the $a_j, b_j$ being any choice of closed paths on the lattice representing a basis of the first homology group.
 We can proceed in the following way: start with closed but not harmonic differentials $\hat{A}_k,\hat{B}_k$ with the periods required by (\ref{harmbasis}). A possible choice for these differentials is shown in Fig. \ref{fig6} where we take $\hat{A}_k$ ($\hat{B}_k$) to be zero on all edges except the ones crossed by the dual lattice cycles $\tilde{b}_k$ (respectively $\tilde{a}_k$). 

Since closed differentials are orthogonal to co-exact differentials it follows from (\ref{hdg}) that closed differentials can be written as the sum of a harmonic differential with the same periods and an exact differential
\bs
      \hat{A}_k&=&A_k+d f^{A}_k \label{decomp}\\
      \hat{B}_k&=&B_k+d f^{B}_k \nn
\es
 applying  $ \delta$ on the these  equations and using (\ref{harm}) and (\ref{deflap}) we obtain 
\bs
  \Delta\, f^{A}_k &=& \delta\,\hat{A}_k \label{eqnum} \\
  \Delta\, f^{B}_k &=& \delta\,\hat{B}_k  \nn
\es

The Laplacian operator on functions (\ref{lap}) is a well defined ${\cal N}\times {\cal N}$ matrix of  rank $({\cal N}-1)$ acting on the functions ${\cal N}$-vector and (\ref{eqnum}) can be solved numerically for reasonably large ${\cal N}$ by fixing the value of the functions  at a lattice vertex. The  solutions can then be  differentiated and subtracted to the $\hat{A}_k,\hat{B}_k$ to obtain a normalized basis of the space of harmonic differentials $\{A_k, B_k :\;  k=1,2\}$. This method is numerically more efficient than the direct solution of the linear system (\ref{harm}).

\section{Holomorphic Differentials and the Period Matrix}\label{section5}

To evaluate a discrete approximation to the period matrix we need to introduce the concept of lattice holomorphic differentials and decompose the  space of harmonic differentials $\{A_k, B_k :\;  k=1,2\}$ obtained in the previous section into the holomorphic and anti-holomorphic sub-spaces. 

In the continuum theory this is accomplished by the projection operators
\be 
P=(1 + i\bigstar)/2 \ \ \text{and} \ \ \bar P=(1 - i\bigstar)/2 \label{proj}
\ee
where  $\bigstar$ is the (continuum) Hodge operator\cite{farkra,hodge}. This operator is an endomorphism on the space of  harmonic differentials and satisfies \mbox{$\bigstar^2=-1$}. Its action on differentials is given by 
\be
     \bigstar\left(f_x(x,y)\, dx + f_y(x,y) dy\right)= -f_y(x,y)\, dx + f_x(x,y) dy 
\ee
and a differential is said to be holomorphic if it is of the form $P\, w$ with $w$ harmonic.
 
We will proceed in a similar way on the discrete theory and define lattice holomorphic differentials  by a projection with a discrete Hodge operator $\star$. 
It proves very difficult to define a discrete Hodge operator acting on lattice differentials, on the finite size lattice, with the same properties of the continuum one. We propose the following definition
\bs
  (\star w)[n|1] &=& -w[\dw(n)|2]\ (h/v)^{-1}  \label{star}\\
  (\star w)[n|2] &=& \ \ w[\lf(n)|1]\ (h/v) \nn .
\es

The action of the discrete Hodge star on the lattice differentials is shown in Fig. \ref{merd}. In this figure an arrow connecting two edges represents that the value of  $\star w$ at the second edge is obtained from the value of differential $w$ at the first edge. Besides the definition above there are three related definitions of the Hodge star, corresponding to $\pi /2$ rotations, that are also shown in Fig. \ref{merd}. Due to the symmetries of the lattice, the four different definitions produce the same numerical results for the period matrix.

The principal merit of these definitions is the way they relate the exterior derivative $d$ with the co-derivative $\delta$. From the definition above and (\ref{der}) and (\ref{del}) it follows that
\bs
  (d\star\! w)[q]&=& - \sum_i \delta \, w[\hat{q}_i]  \label{ddstar1} \\
  (\delta \star\! w)[n]&=&\left\{
              \begin{array}{cl}
                   \alpha(n)  & \text{   if }  n \in \{n_1,n_2,n_3,n_4\}\\ 
                    d\, w[\widetilde{\dw}(\lf(n))] & \text{   otherwise }
              \end{array}   \right. \label{ddstar2}
\es
where in the first equation the sum goes over $i$=1,2 for octagonal faces and is $i$=1 for squared faces. The $n_k$ are the four lattice vertices where ${\lf}(\dw(n_k))\neq{\dw}(\lf(n_k))$. Let $n_1$ and $n_2$ be on the same octagon and the two remaining points on the other octagon then the $\alpha(n_k)$ satisfy
\bs
    \alpha(n_1)+\alpha(n_2)&=& d\, w[\widetilde{\dw}(\lf(n_1))]  \\
    \alpha(n_3)+\alpha(n_4)&=& d\, w[\widetilde{\dw}(\lf(n_3))] \nn   .
\es

The appearance of the $\alpha$ terms in (\ref{ddstar2}) prevents the discrete Hodge operator from being an endomorphism on the harmonic lattice differentials. These terms originate from the fact that the lattice has more vertices than faces: the requirement that a differential $w$ is closed is not sufficient to ensure that the corresponding  $\star w$ is co-closed.  For the dual lattice, where the number of faces is larger than the number of vertices, the opposite happens and the $\alpha$ terms appear in (\ref{ddstar1}) instead of (\ref{ddstar2}).

Unlike the continuum Hodge star, our lattice definition does not satisfy $\star^2=- 1$ but rather
\bs
  \star^2 w[n|1]&=& -w[\lf(\dw(n))|1] \label{sqstar}  \\
  \star^2 w[n|2]&=& -w[\dw(\lf(n))|2]  \nn
\es
there is a displacement of the differential and mixing at the octagons. 

Notice that for a toroidal square lattice, where all faces are squares and  \mbox{${\lf}(\dw(n))={\dw}(\lf(n))$} for all vertices $n$, there are no $\alpha(n)$ terms on the equivalent of equations (\ref{ddstar1}) and (\ref{ddstar2}). The lattice Hodge star (\ref{star}) on a torus is an endomorphism on the space of harmonic differentials.

 Lets proceed to evaluate the torus modular parameter $\tau$ and compare with the know result (\ref{tau}). Consider a  toroidal square lattice with $M$ rows and $N$ columns. Let $\{a,b\}$ be a basis of the first homology group with $a$ being a horizontal loop and $b$ a  vertical loop. This lattice has two independent  normalized harmonic differentials $A$ and $B$ with components
\be
 \begin{array}{ll}   
   A[n][1]=1/N, &  A[n][2]=0 \\
   B[n][1]=0, & B[n][2]=1/M\\
 \end{array}  
\ee
for all vertices $n$. The only holomorphic differential on the lattice is $\Gamma=A+i\star A$, in components,
\bs
   \Gamma[n][1]=\frac{1}{N},\ \  \Gamma[n][2]=i\frac{h}{v}\frac{1}{N}
\es
for all vertices $n$.
The modular parameter is the integral of this differential along a loop of the homology class of $b$ and we obtain
\be
   \tau=\int_{b} (A+i\star A)=i\frac{h}{v}\frac{M}{N}
\ee
that reproduces the well known result (\ref{tau}) if we take for the $h/v$ ratio
\be
     h/v = \left\{
              \begin{array}{cl}
               \cosh{2 K_h^c}/\cosh{2 K_v^c} &  \text{    for the critical Ising Model } \label{pedro}\\ 
                z_h/z_v                     &  \text{    for the Close-Packed Dimers} 
             \end{array}   \right.    .  
\ee

For the Ising model this ratio corresponds to the ratio of the diverging correlation lengths along the two lattice directions\cite{pearce}, the relevant geometrical quantity at criticality. We will use the same expression  for the genus two case, however in this case the lattice Hodge operator  (\ref{star})is not an exact endomorphism on the harmonic lattice differentials, due to the $\alpha$ terms that appear on the four $n_k$ vertices. 

We will take (\ref{star}) as the definition of our lattice Hodge operator and  proceed to evaluate a basis for the space of differentials of the form \mbox{$\{\Gamma_k=(1+i\star)H_k : k=1\ldots,g\}$} where the $H_k$ are harmonic differentials, satisfying the normalization conditions 
\bs
    \int_{a_k} (H_l+i\star H_l) &=& \delta_{kl} \label{holo} \\
     \int_{b_k}  (H_l+i\star H_l) &=& \Omega_{kl} \label{period}
\es

Equation (\ref{holo}) is a set of $2g$ real constraints on $H_k$ from which we can determine the coefficients of  $H_k$ in the basis  $\{A_k, B_k\}$ and equation (\ref{period}) gives our approximation to the lattice period matrix.

Since \mbox{$\{\Gamma_k=(1+i\star)H_k : k=1\ldots,g\}$}, seen as a complex vector space, is not invariant under  $P=(1+i\star )/2$ we are not allowed to call these lattice differentials holomorphic in a strict sense. However a detailed numerical study, discussed on the next section, will show that the period matrix obtained by this procedure reproduces to excellent precision the period matrices of equation (\ref{resultati}).

\section{Period matrices and determinant ratios} \label{section6}

Following the procedure described in the two previous sections we can  evaluate numerically the period matrix for lattices with different aspect ratios $(m_1,m_2,k,n_1,n_2)$ and coupling constants. Table \ref{table6} shows the period matrix for a particular lattice  aspect ratio $(m_1,m_2,k,n_1,n_2)$ and various lattice sizes $(M_1,M_2,K,N_1,N_2)= (m_1,m_2,k,n_1,n_2)\,L$, with increasing $L$. The period matrix elements converge  in a smooth way with the number of lattice vertices $\cal N$ and  the resulting matrix is a purely imaginary, symmetric and positive definite matrix for all lattice sizes.

The three different set of values shown in  table \ref{table6} correspond to three different choices of the first homology group basis. The first set of values, labeled as A, corresponds to the choice of homology basis  $a_i, b_i$ shown in Fig. \ref{fig1}; the set labeled as B  corresponds to a choice where the $a_i$ and $b_i$ loops are interchanged from  Fig. \ref{fig1} and the set C corresponds to the choice where the top and bottom of Fig. \ref{fig1} are interchanged.
The different period matrices obtained for each choice of the homology basis  are related by a modular transformation. In  table \ref{table7} the  theta function ratios corresponding to each choice are shown  to be related by  permutations. While the A and C sets of values agree at finite size, the  period matrix corresponding to B gives  numerical values different from the other two.  This fact is related with the $\alpha$ terms in  equations (\ref{ddstar1}) and (\ref{ddstar2}) that keep the lattice Hodge operator from being an endomorphism on lattice harmonic differentials.

While at finite size, modular invariance is  slightly broken this does not seem to be the case in the continuum limit.  In  table \ref{table7} we see that the theta function ratios extrapolated to the ${\cal N}\rightarrow \infty$ limit, agree within a $10^{-2}$ error. 
 Fig. \ref{fig69} and \ref{fig70}  show the theta function ratios corresponding to the two different modular choices A and B,  plotted as a function of number of lattice vertices $\cal N$, together with the determinant ratios. 
In these figures we observe that theta ratios for the period matrices A and B seem to converge to a common value and that, for large enough lattice size, the values of the nine non-vanishing determinant ratios are bounded by two sets of the theta ratios. 
Qualitatively the pattern observed in these plots leads us to believe that the procedure described on the previous section to evaluate holomorphic differentials and the period matrix has a well defined continuum limit.  

A quantitative analysis is given in tables \ref{table8}  and \ref{table9} where the period matrices evaluated by the procedure above (using the choice A for the homology basis) are compared with the period matrices of equation (\ref{resultati}), obtained from the ratios of adjacency matrices determinants by numerical fit. There is a remarkable agreement between the two sets of values.
The theta function ratios  corresponding to the two period matrices agree with a precision of $10^{-3}$ or better. 

The small numerical differences are associated with the difficulty in extrapolating the finite size values of the period matrix to the continuum limit. The extrapolated values given in tables \ref{table8}, \ref{table9}, \ref{table6} and \ref{table7} are obtained by fitting the size  dependence of the period matrix entries with a quadratic polynomial in  $\cal N$. 

As in the toroidal case, the vertical and horizontal coupling constants provide different weighting to the vertical and horizontal directions.
The period matrix for a lattice characterized by the aspect ratio $(m_1,m_2,k,n_1,n_2)$ and parameter $h/v$ is equivalent to the period matrix of a lattice characterized by the aspect ratio $(\alpha m_1,\alpha m_2,\beta k,\beta n_1,\beta n_2)$ and a re-scaled parameter $\beta h/\alpha v$. This property is exemplified in the first two sets of values in tables \ref{table8}  and \ref{table9}  for both the critical Ising model and Dimers. The parameter $h/v$ can therefore be absorbed into a re-scaling of the overall vertical/horizontal aspect ratio of the lattice.

\section{Conclusions} \label{section7}

In this paper we used the Kasteleyn formalism to study Dimers and the critical Ising models on a genus two lattice. It is the first time that such a study has been done in higher genus lattices. We found  that the determinants of the Kasteleyn adjacency matrices converge (\ref{resultati}) to a common term times theta functions of half-integer characteristic. This result generalizes a thirty year old observation on the asymptotics of Pfaffians of adjacency matrices on the torus\cite{ferd,ferdfish,nashoco} and gives a new meaning to this convergence.

 For the critical Ising model, the dependence of the determinants of adjacency matrices on the Kasteleyn orientations is exactly the same as the dependence of the determinant of the Dirac operator, of the corresponding conformal field theory, on the spin structures of the  Riemann surface; both are given in terms of theta functions of half integer characteristics. There is therefore  an one to one correspondence between the $4^g$ Kasteleyn clockwise odd orientations of the lattice and the $4^g$ spin structures of the continuum limit Riemann surface. 

These observations elevate the Kasteleyn formalism from a combinatorial trick to obtain the partition function to a discrete formulation of some of the model conformal field theory  properties. 
The relation between the lattice determinants and the zeta regulated determinants of the conformal field theory is reminiscent of the  Ray-Singer theorem\cite{raysin,muller} that relates  determinants of the analytical Laplacians on a surface with the determinants of combinatorial Laplacians on a triangulation of that surface.

In the context of conformal field theory  the period matrix is usually seen as a free parameter and, except on the torus, is not related to the thermodynamic limit of a lattice or a set of particular lattices. We have shown that the period matrix, characterizing the continuum limit of a lattice, can be understood in terms of a lattice discrete holomorphic structure. 

An important difference between the genus two case (\ref{z2}) and the torus  (\ref{z1}) is that the determinants corresponding to the odd characteristic theta functions do not vanish at finite size and only converge to zero on the $\cal N \rightarrow \infty$ limit. 
It interesting to notice that on the toroidal square lattice a completely satisfactory definition of discrete holomorphy at finite size can be given, a fact that is probably related with the finite size vanishing of the odd characteristic determinant. 

The results we obtained for the  genus two case can be readily generalized for arbitrary genus. Higher genus lattices can be constructed by pasting two or more of the lattices of Fig. \ref{fig1}, as shown in  Fig. \ref{fig8}. The construction of the adjacency matrices and period matrices can also be done in a similar way to the one described in this paper for genus two.
 
There are many ways of choosing the boundary conditions of a square lattice to obtain a higher genus lattice. An important requirement to reproduce the results we obtained is that the boundary conditions do not destroy the  distinction between vertical and horizontal edges. With the  boundary conditions suggested in Fig. \ref{fig8}, the genus $g$ lattice has a total of $3g-1$ integer sizes, respectively $(M_1,\ldots,M_g,K_1,\ldots,K_{g-1},N_1,\ldots,N_g)$. These account for $3g-2$ independent aspect ratios, the $h/v$ parameter been absorbed into a combination of these ratios. These  $3g-2$ rational parameters cannot cover, or even approximate, all the complex structures on a genus $g$ Riemann surface, that are parameterized by $3g-3$ complex parameters. This may be possible by generalizing the genus two lattice of Fig. \ref{fig1} from locally square to  locally triangular and allowing for different couplings constants in each sub-lattice.

\bigskip
\centerline{{\bf Acknowledgments}}
\bigskip

 This work is partially supported by the Funda\c{c}\~ao para a Ci\^encia e Tecnologia (Portugal) under Grant BD 11503 97 and by the National Science Foundation (USA) under Grant No. DMR-0073058.
Both authors profited from useful discussions with P. Pearce and C. Mercat.  B.M.M. is pleased to thank M. Kashiwara and T. Miwa for hospitality at the
Research Institute of Mathematical Sciences where part of this work was
completed. R.C-S. thanks Rui Sousa for his help with the numerical computations.

\begin{appendix}

\section{Kasteleyn combinatorics} \label{append1}

\begin{figure}[h]
\vskip 2cm
\epsfig{figure=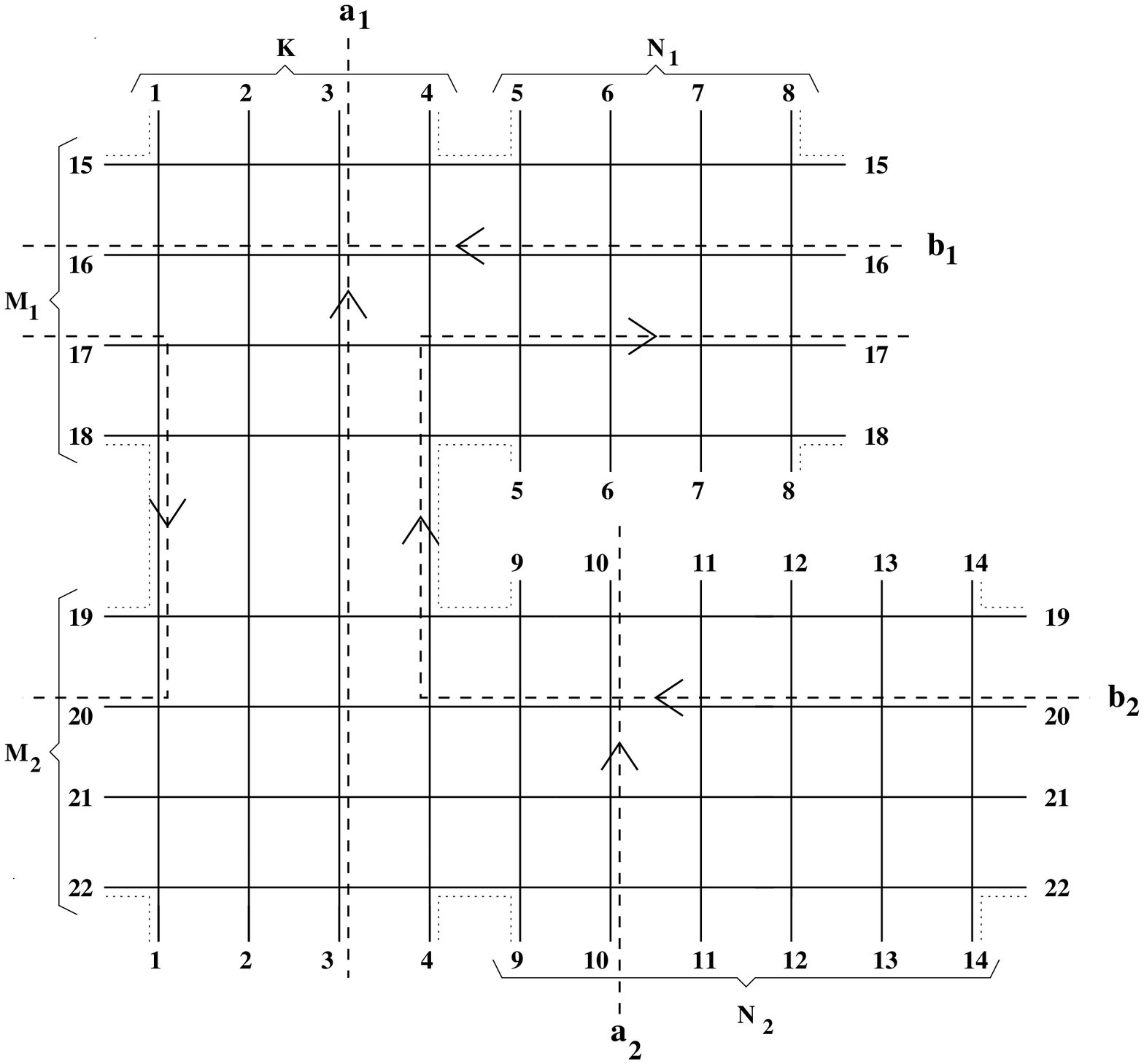,width=15cm}
\vskip 1cm
\caption{The genus two lattice with the boundary identifications given by the integers from 1 to 22. The $a_i,b_i$ cycles form a basis of the first homology group and should be seen as drawn over the lattice edges. All lattice faces are squared except for two octagons whose edges are marked with the doted line.}
\label{fig1}
\end{figure}

In this appendix we derive  (\ref{zzz}), the Pfaffian expansion of the  partition function. We will show that from the Pfaffians of the sixteen Kasteleyn adjacency matrices $\Pf (A_i)$ we can obtain the partition functions of Ising and Dimers for sixteen different choices of boundary conditions on the lattice of Fig. \ref{fig1}. 
 Let $Z(0000)$ denote the partition function of Dimers or Ising on the lattice  of Fig. \ref{fig1}, then
\be
     Z_i=Z(n_{\tilde a_1},n_{\tilde b_1},n_{\tilde a_2},n_{\tilde b_2})
\ee 
with $i=16-8n_{\tilde a_1}- 4n_{\tilde b_1}-2n_{\tilde a_2}-n_{\tilde b_2}$ and  $n_x=0,1$ for $x=n_{\tilde a_1},n_{\tilde b_1},n_{\tilde a_2},n_{\tilde b_2}$ represent the partition functions of the  same model with coupling constants or dimer weights multiplied by $-1$ along a choice of the $n_x=1$ representatives of the first homology group. This is a labeling similar to the one introduced for adjacency matrices in  (\ref{defori}). 

We will show that there is a matrix  $b_{ij}$ such that
\be
     Z_i= \sum_{j=1}^{4^g}\; b_{ij} \;\Pf ( A_j) \ \ \text{  and } \ \ \Pf ( A_i)= \sum_{j=1}^{4^g}\; b_{ij}\; Z_j  \label{zi} 
\ee 
The property that $b^2=1$ was first noticed for the torus by  Fradkin and Shteingradt\cite{fradkin}. This places the partition functions and the Pfaffians of the adjacency matrices on equal footing.
For both models the partition function $Z(n_{\tilde a_1},n_{\tilde b_1},n_{\tilde a_2},n_{\tilde b_2})$ is independent of the choice of  representatives of the cycles $a_k,b_k$. In the Ising model it is well known that a deformation of a disorder loop does not alter the partition function since a disorder loop that crosses a vertex corresponds to an interchange of the up and down spin at that vertex. For the Dimer problem the invariance of the partition function   $Z(n_{\tilde a_1},n_{\tilde b_1},n_{\tilde a_2},n_{\tilde b_2})$ follows from (\ref{zi}) and from the invariance of the determinants of adjacency matrices under disorder loop deformation.

To keep the paper self contained we start by  giving a short overview of the Kasteleyn formalism in higher genus lattices following a combination of references\cite{kast3,mccoy,russ,tesl}. 
By a genus $g$ lattice we mean a graph, not necessarily regular, that can be drawn without superposition of edges only in a surface of genus $g$ or higher. 
The  Ising model decorated lattice,  Fig. \ref{fig3}, has crossing edges and cannot be  considered a genus two lattice. To apply the results of this appendix to the Ising model it is preferable to consider the Fisher six vertices decoration\cite{fisher} instead of the Kasteleyn four vertices decoration, see Fig. \ref{fig99}. The Fisher decorated version of our lattice can be embedded on a genus two surface without superposition of edges. Determinants of adjacency matrices of the two lattices are equal\cite{mccoy}.  We will use the Fisher lattice on the formal discussion and the Kasteleyn lattice for actual numerical calculations.

Recall the definition of the Pfaffian of anti-symmetric matrix, equation (\ref{pff}) 
\be
    \Pf  (A) = \frac{1}{2^{N/2}(N/2)!} \sum_p \epsilon{\scriptsize\left(\begin{array}{cccc}1&2&\cdots&N\\p_1&p_2&\cdots&p_N\end{array}\right)} \;A_{p_{1}p_{2}}A_{p_{3}p_{4}}\cdots A_{p_{N-1}p_{N}}
\ee 
where $p$ is a permutation of the integers from 1 to N and $\epsilon_p=\pm 1$ for even and odd permutations respectively .
The normalization factor $2^{N/2}(N/2)!$ accounts for the fact that permutations  differing by the order of the indices in a  $A_{p_{i}p_{i+1}}$ factor or by the interchange of two such factors give equal contributions to the Pfaffian. We will call two permutations related in such a way equivalent permutations. 

 A non zero Pfaffian term $A_{p_{1}p_{2}}A_{p_{3}p_{4}}\cdots A_{p_{N-1}p_{N}}$ corresponds to a choice of edges $C_p=\{(p_1,p_2),\ldots,(p_{N-1},p_N) \}$ where no edges share a common vertex and all vertices are included;  $(i,j)$ representing the edge between vertices $i$ and $j$. This choice of edges is a dimer configuration with weight equal, in absolute value, to the Pfaffian term.  Dimer configurations are therefore  in one to one correspondence with sets of equivalent permutations with non zero Pfaffian term.

To express the dimer partition function  in terms of Pfaffians of adjacency matrices we need to choose edge orientations such that all the terms in the Pfaffian have the same relative sign or a linear combinations of Pfaffians with this property. 
The relative sign between  Pfaffian terms corresponding to different permutations $p$ and $p'$  can studied  by considering  their product
\be
  \epsilon{\scriptsize\left(\begin{array}{cccc}1&2&\cdots&N\\p_1&p_2&\cdots&p_N\end{array}\right)}\epsilon{\scriptsize\left(\begin{array}{cccc}1&2&\cdots&N\\p'_1&p'_2&\cdots&p'_N\end{array}\right)} \;A_{p_{1}p_{2}}\cdots A_{p_{N-1}p_{N}} \;A'_{p'_{1}p'_{2}}\cdots A'_{p'_{N-1}p'_{N}} \label{pro}
\ee 
where $A$ and $A'$ represent the same adjacency matrix, the prime being introduced for convenience. This product has a simple graphical interpretation: it corresponds to the superposition diagram of the two dimer configurations $C_p$ and $C_{p'}$. The superposition diagram is the set of double edges and  even length cycles obtained by drawing both dimer configurations on the lattice, see Fig. \ref{fig9} for an example. The even length cycles have alternating edges belonging to each one of the two  dimer configurations and are  called transition cycles.

The product  (\ref{pro}) can always be rewritten in a form that resembles the superposition diagram\cite{kast3}
\be
\epsilon{\scriptsize\left(\begin{array}{cccc|cc|c|c}i_1&i_2&\cdots &i_q&i_{q+1}&i_{q+2} &\cdots&\cdots  \\i_2&i_3&\cdots &i_1&i_{q+2}&i_{q+1}  &\cdots&\cdots  \end{array}\right)}(A_{i_{1}i_{2}}A'_{i_{2}i_{3}}A_{i_{3}i_{4}}A'_{i_{4}i_{5}}\cdots A'_{i_{p}i_{1}})( A_{i_{q+1}i_{q+2}} A'_{i_{q+2}i_{q+1}})(\cdots)\cdots     
\ee
where we used the fact that the sign of a product of permutations is the product of the sign of each permutation and  we decomposed the overall permutation into  cyclic permutations of even length. We also chose equivalent permutations to $p$ and $p'$ and rearranged the matrix elements in such a way that the terms corresponding to the double edges and the even length cycles of the superposition diagram are singled out. The equation above is a possible example where one cycle of 2p length and a double edge are shown. 

{}From this formula we can read out the relative sign of two Pfaffian terms. The sign of the overall permutation is $(-1)^s$ where $s$ is the total number of even length cycles, including double edges, in the superposition diagram. {}From the double edges $A_{ij}A'_{ji}$  we get an additional minus sign since $A$ is anti-symmetric, while each transition cycle contributes with a sign given by $(-1)^p$ where $p$ is the number of edges in that cycle oriented, say,  in the clockwise direction. 

Following the notation of reference\cite{russ} let $C\tri C'$ denote the set of transition cycles resulting from the overlap of the dimer configurations  $C$ and $C'$ with the double edges removed. 
Then the above discussion motivates the following result due to Kasteleyn\cite{kast3}

\begin{theo}
Let $\text{sign}(C)$ be the sign of the terms in the Pfaffian corresponding to a given dimer configuration $C$. Then for any two dimer configurations $C$ and $C'$ we have that  

\be
 \text{sign}(C) \text{sign}(C')=  \; \prod_{\gamma} (-1)^{p(\gamma)+1} \label{sign}
\ee
where the product runs over all transition cycles $\gamma$ in $C\tri C'$ and $p(\gamma)$ is the number of clockwise oriented edges in the cycle $\gamma$.
\end{theo}

We will call $\tri$ the overlap operator. If dimer configurations are seen as sets of edges then  $$C_1\tri C_2= (C_1\cup C_2)\backslash (C_1\cap C_2)$$ is the symmetric difference of the two sets.
Since superpositions of configurations will play a major role in the  following discussion it is worth to collect some properties of the overlap operator $\tri$. The following properties follow directly from the definition, the $C_i$ being dimer configurations or more generally sets of edges 

\bs
  C_1\tri C_2&=& C_2\tri C_1  \label{prop}\\
   C_1\tri C_1&=& \emptyset \nn   \\
    C_1\tri  \emptyset&=& C_1 \nn \\
   C_1 \tri (C_2 \tri C_3) &=& (C_1 \tri C_2) \tri C_3 \nn 
\es

We can use this overlap operator to classify the dimer configurations, and the corresponding Pfaffian terms, into  equivalence classes or types. We start by choosing a  standard dimer configuration $C_0$, the standard configuration for our lattice being shown in Fig. \ref{fig9}.
 Let  $a_i, b_i$ with $i=1,\ldots,g$ be a canonical basis of the first homology group on the lattice then we say that $C$ is of type $T(n_{b_1}, n_{a_1},\ldots, n_{b_g}, n_{a_g})$ if there are $n_x$  topologically non-trivial cycles of kind $x=a_i,b_i$  present in the overlap  $C_0\tri C$. An example for our genus two lattice is show  in Fig.  \ref{fig9} where a dimer configuration $C$ is seen to be of type $T(0102)$.

We are then classifying $C$ by the homology class of the overlap $C_0\tri C$, two dimer configurations $C$ and $C'$ being of the same type if  $(C_0\tri C)=(C_0\tri C')\tri X$ where $X$ is a boundary, that is, a set of closed cycles with trivial homology. 
Since the overlap operator $\tri$ eliminates double edges it can only distinguish homology classes modulo two. For instance in Fig.  \ref{fig10} we show  how a $T(2001)$ superposition can be obtained from a $T(0001)$ superposition by overlapping a homologically trivial boundary $X$. We will therefore take $n_x=e,o$ where were $e$ stands for even and $o$ stands for odd.
 
The relevance of these equivalence classes is that the relative sign of a configuration C to the standard configuration $C_0$ for a clockwise odd edge orientation depends only on the equivalence class of C. 
Clockwise odd edge orientations were defined in section \ref{section2} to be orientations such that every lattice face has an odd number of clockwise oriented edges along its boundary (the clockwise orientation is a convention, anti-clockwise odd edge orientations would work equally well). These edge orientations have the following property, found by Kasteleyn\cite{kast1} 

\begin{theo}
For an edge orientation such that all lattice faces have an odd number of clockwise oriented edges we have that  $ \text{sign}(C) \text{sign}(C')=1$ for any pair of dimer configurations $C$ and $C'$ such that $C\tri C'$ has a trivial homology.
\end{theo}   

A complete proof of this theorem can be found in references\cite{kast3,mccoy} but the general idea is very simple: if  $C\tri C'$ has a trivial homology its  transition  cycles  can be built by the successively overlap of elementary lattice faces with the $\tri$ operator. It can then be shown that any such  transition cycle  will have clockwise odd parity and contribute with a  minus sign in the Pfaffian term that  cancels the minus sign of the corresponding cyclic permutation.

We can now state the main result of the higher genus Kasteleyn formalism, whose proof is so simple that we present it here

\begin{theo}
The relative sign of a dimer configuration $C$ to the standard configuration $C_0$ in a clockwise odd orientation depends only on the homology class modulo two of the set of transition cycles $C_0\tri C$.
\end{theo}
\textbf{Proof:} Let $C$ and $C'$ be two dimer configurations such that   $C_0\tri C$ and $C_0\tri C'$ belong to the same homology class. Then by definition we must have $(C_0\tri C)=(C_0\tri C')\tri X$ were $X$ is a set of homologically trivial cycles. Using (\ref{prop}) we find that $X= C\tri C'$ and  it follows from theorem 2 that, in  a clockwise odd orientation,we have $ \text{\it sign}(C) \text{\it sign}(C')=1$. $\Box$
\\

In a genus $g$ lattice  there are $4^g$ classes of dimer configurations, the number of elements of the  homology group $H_1(G,Z_2)$,  that in general will have different  signs in a given clockwise odd orientation.  Conversely there are also $4^g$ inequivalent Kasteleyn orientations, the number of elements of the  cohomology group $H^1(G,Z_2)$ and it is possible to find a linear combination of the corresponding   $4^g$ Pfaffians of adjacency matrices such that all dimer configurations have the same overall sign.

More precisely: two edge orientations are said to be equivalent if they are related by a sequence of local transformations in which we reverse the edge orientation of all the edges coming to a given vertex. It is clear that each local transformation is equivalent to the multiplication of the i-th column and row of $A$ by $-1$ and will not affect the parity $p(\gamma)$ of the transition cycles $\gamma$ or the absolute value of $\det A$. We then have the following result

\begin{theo}
 In a genus $g$ lattice there are $4^g$ inequivalent Kasteleyn edge orientations.
\end{theo}

For our  genus two lattice it is easy to see that there are only  sixteen inequivalent Kasteleyn orientations.  Consider the lattice as drawn in Fig. \ref{fig11}. We can choose the edge orientation on all the edges of a spanning tree of the lattice (heavy line) using  the freedom given by the local equivalence transformations. Then the condition that the elementary faces must be clockwise odd will fix the orientation on most edges (the dashed line edges) but there will be 4 edges (rectangular edges in the figure) in which the orientation is undetermined. The sixteen different choices of orientations in these edges will correspond to sixteen inequivalent Kasteleyn orientations.
Using a similar reasoning the reader can convince her/himself that the theorem is true for any lattice.

In section 2 we constructed the sixteen edge orientations for our genus two lattice by starting from an initial orientation and introducing disorder loops along the $\tilde a_i,\tilde b_i$ cycles. This procedures makes it clear the connection between the edge orientations and the the  cohomology group $H^1(G,Z_2)$.

In order to obtain the linear combination of Pfaffians that yields the partition function  we still need to determine the relative sign that the different dimer configuration types, $T(n_{b_1}n_{a_1}n_{b_2}n_{a_2})$ with $n_x=e,o$, take in  the various clockwise odd configurations, $A(m_{\tilde a_1}m_{\tilde b_1}m_{\tilde a_2}m_{\tilde b_2})$ with $m_{\tilde x}=0,1$.

\begin{figure}[h]
\vskip 3cm
\epsfig{figure=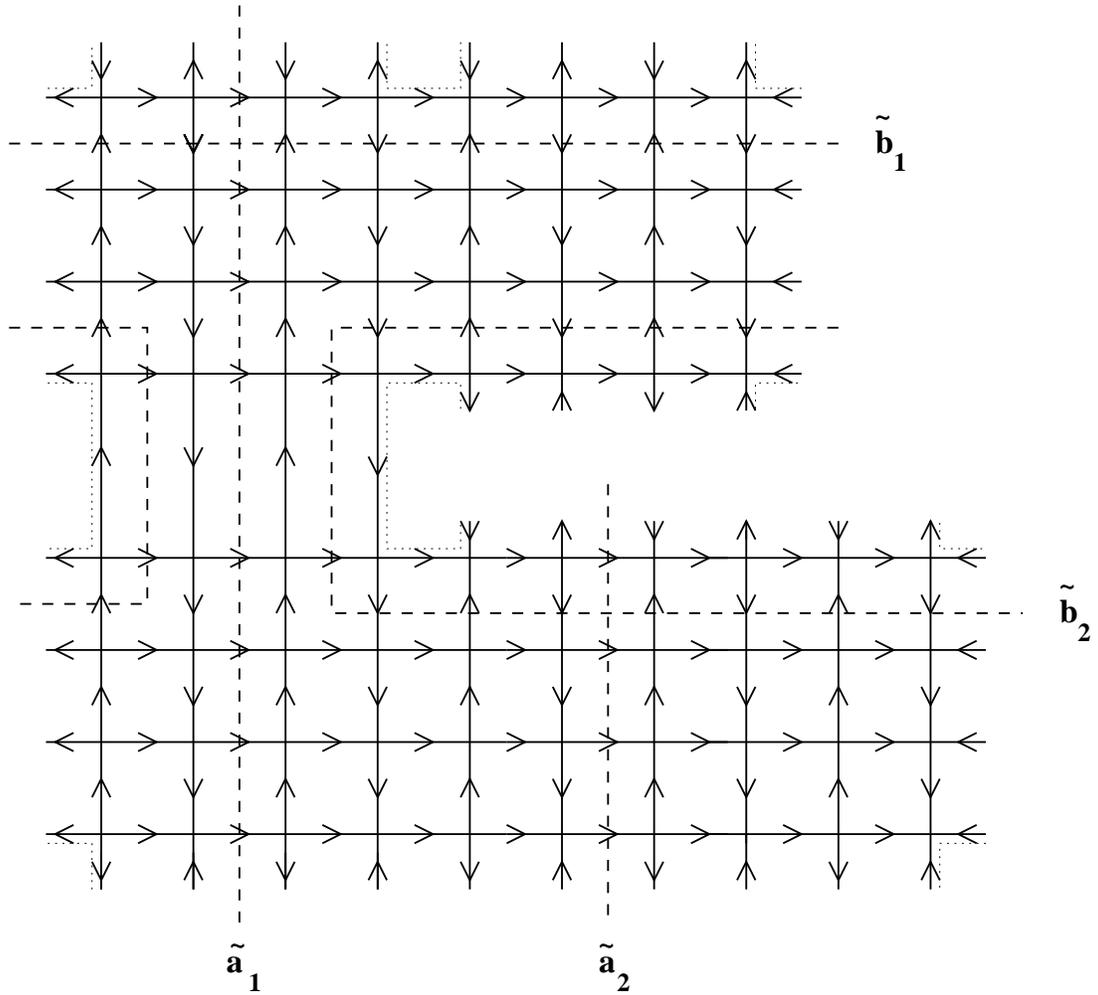,width=15cm}
\vskip 1cm
\caption{The $A(0000)$ clockwise odd orientation of the original lattice. The $\tilde a_i,\tilde b_i$ cycles form a basis of the first homology group drawn over the dual lattice edges. The remaining clockwise odd orientations can be obtained from this one by inverting the orientations of the edges crossed by a given choice of the  $\tilde a_i,\tilde b_i$ cycles.}
\label{fig2}
\end{figure}

\begin{figure}[h]
\vskip 3cm
\epsfig{figure=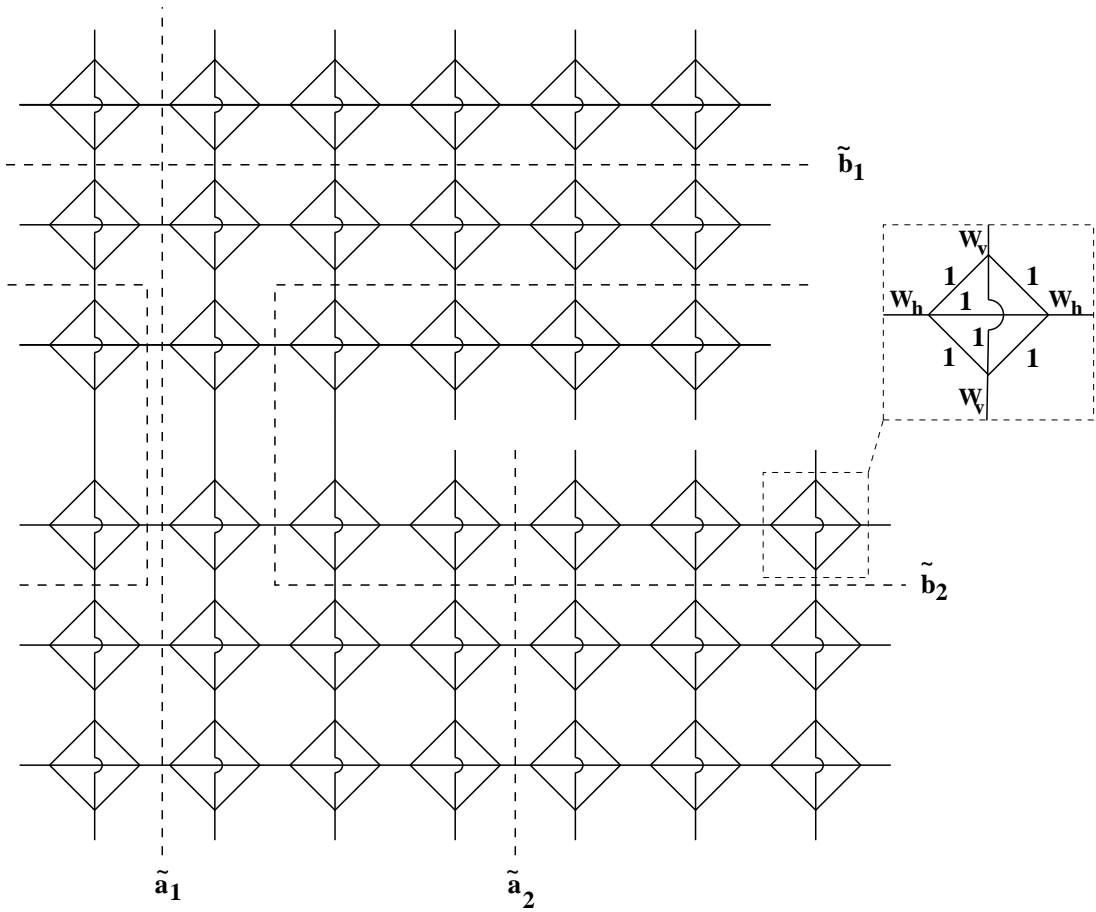,width=15cm}
\vskip 1cm
\caption{The Ising decorated lattice, the dimer weights of each edge type are shown in the box.}
\label{fig3}
\end{figure}

\begin{figure}[h]
\vskip 3cm
\epsfig{figure=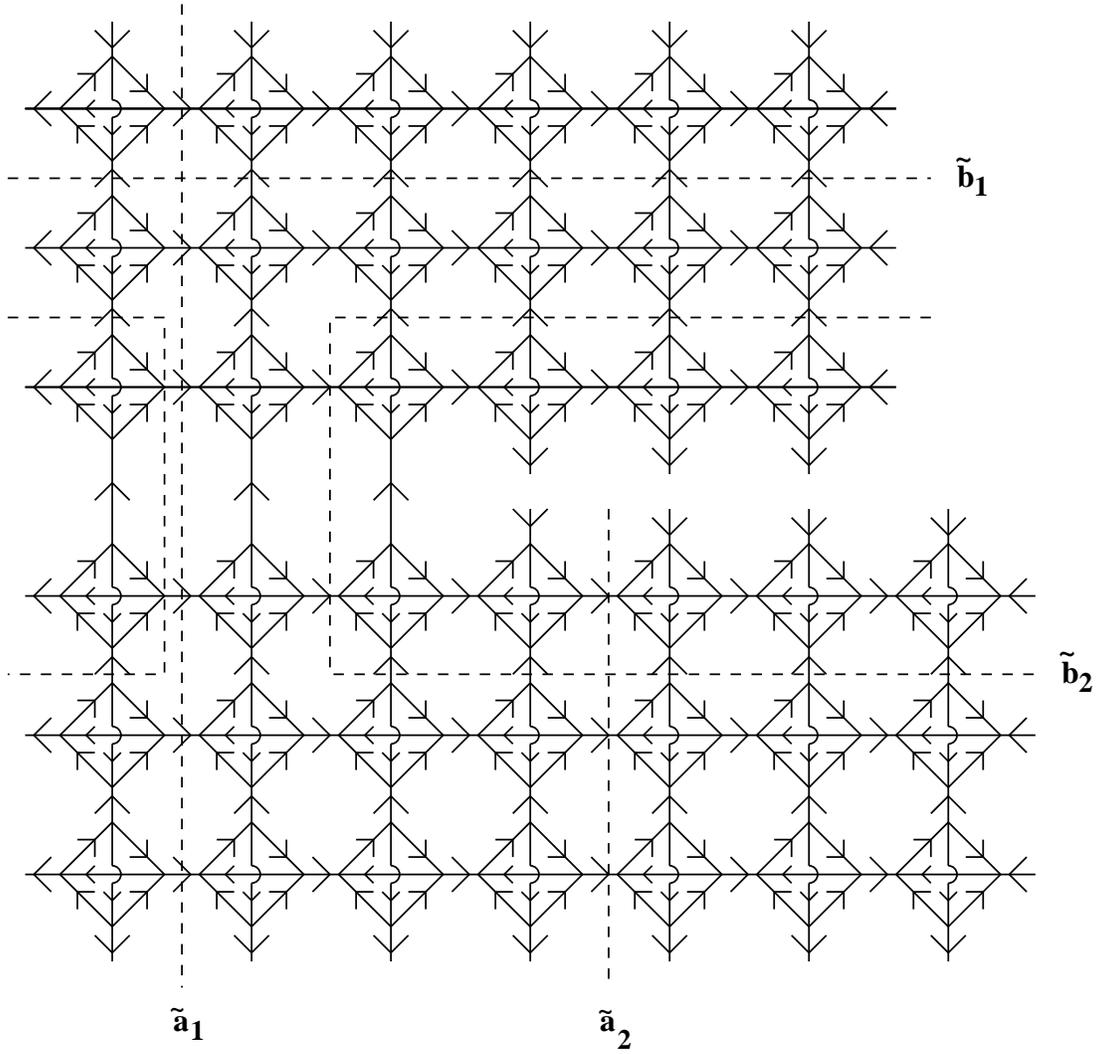,width=15cm}
\vskip 1cm
\caption{The $A(0000)$ clockwise odd orientation of the Ising decorated lattice. The remaining clockwise odd orientations can be obtained from this one by inverting the orientations of the edges crossed by a given choice of the  $\tilde a_i,\tilde b_i$ cycles.}
\label{fig4}
\end{figure}

\begin{figure}[h]
\vskip 3cm
\epsfig{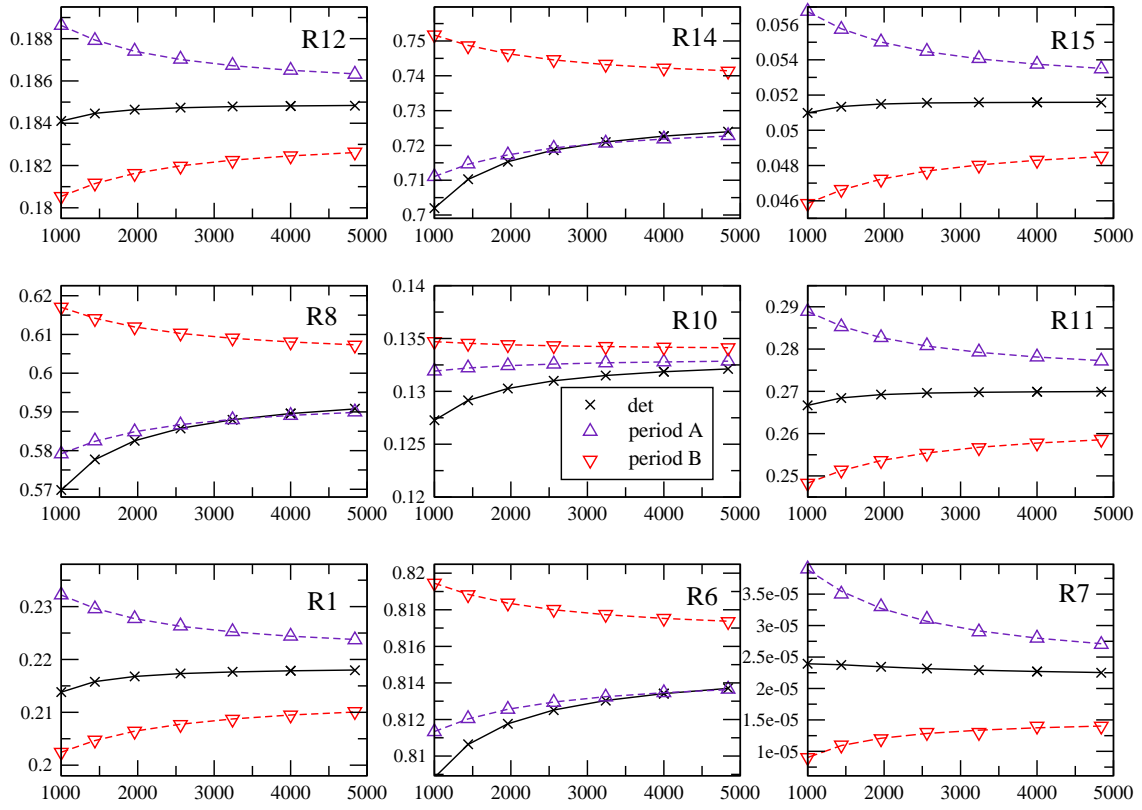}
\vskip 1cm
\caption{The continuum limit for DIMERS: the nine non-vanishing ratios of determinants, defined as $R_i= \det A_i/\det A_{16}$, and the corresponding ratios of theta functions are shown as a function of the number of lattice vertices $\cal N$,  for the lattice  with aspect ratio ($m_1$,$m_2$,k,$n_1$,$n_2$)=(4,2,2,6,2) and  $z_h$=0.780 $z_v$=0.560. The solid curve is the plot of the determinant ratios  and the dashed curves are the plots of ratios of theta functions, with period matrices evaluated for the A and B choices of the homology basis (see section \ref{section6} for discussion). The curves are fits with a polynomial on $1/\cal N$ to the values obtained for different lattice sizes.}
\label{fig69}
\end{figure}

\begin{figure}[h]
\vskip 3cm
\epsfig{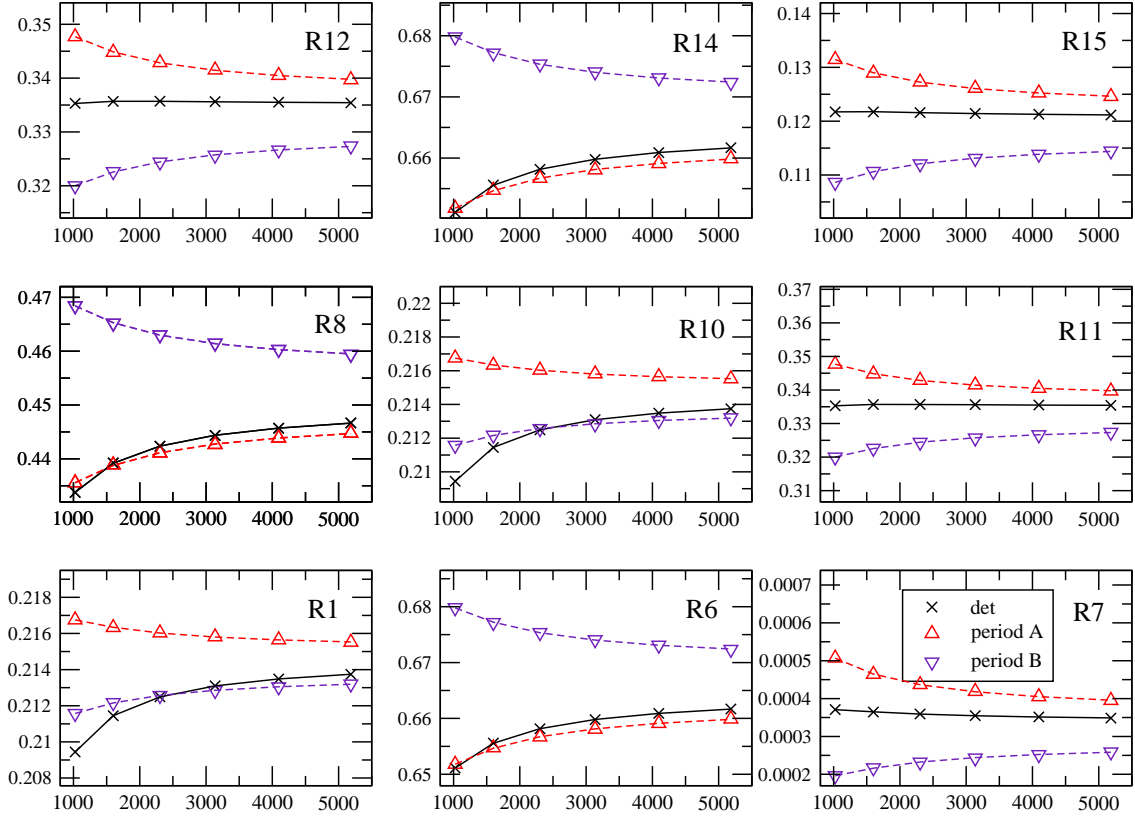}
\vskip 1cm
\caption{The continuum limit for ISING: the nine non-vanishing ratios of determinants and ratios of theta functions are shown, as a function of the number of lattice vertices $\cal N$, for the lattice  with  aspect ratio ($m_1$,$m_2$,k,$n_1$,$n_2$)=(2,2,2,2,2)  and $w_h$=0.537  $w_v$=0.301. Some of the ratios are equal due to the high symmetry of the lattice.}
\label{fig70}
\end{figure}

\begin{figure}[h]
\vskip 3cm
\epsfig{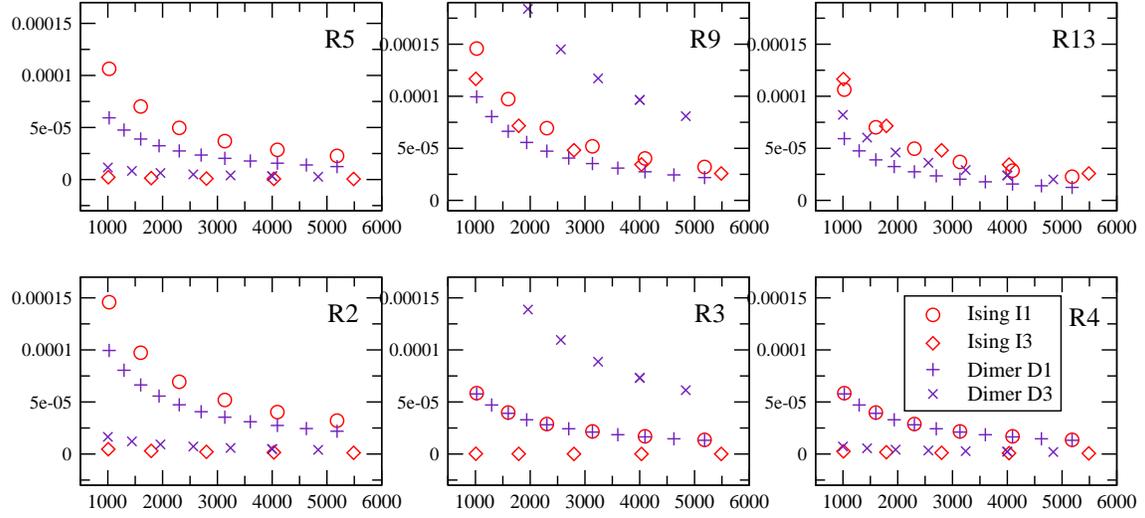}
\vskip 1cm
\caption{The six vanishing ratios of determinants for four different lattices. For comparison Ising determinant ratios are shown squared. Coupling constants and aspect ratios for the lattices I1, I3, P1 and P3 are given in tables \ref{table8} and \ref{table9}.}
\label{fig68}
\end{figure}

\begin{figure}[h]
\vskip 3cm
\epsfig{figure=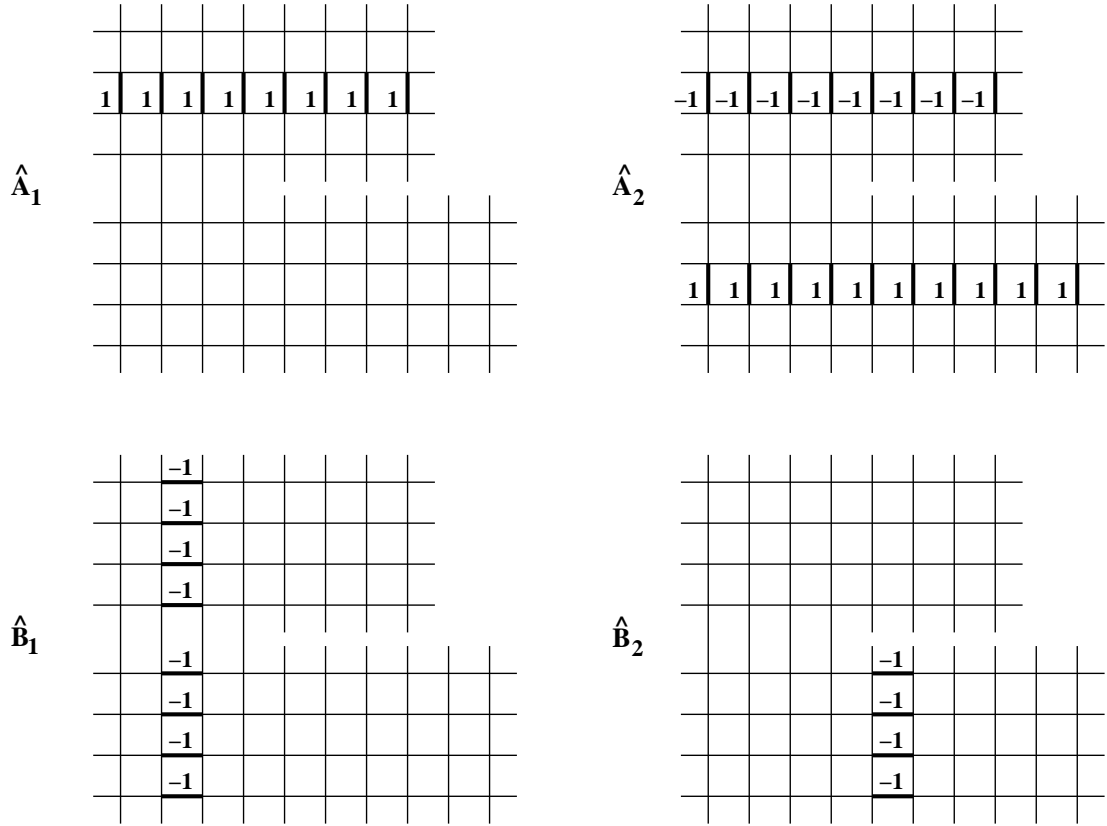,width=15cm}
\vskip 1cm
\caption{The closed differentials $\hat A_k,\hat B_k$. The differentials are zero on all edges except the ones crossed by the respective $\tilde b_k$, $\tilde a_k$ loop. These edges are shown in bold on the figure, together with the value that the differential takes at that edge. For convenience we use a $\tilde a_2$ loop different but homologically equivalent to the one shown in Fig. \ref{fig1}.}
\label{fig6}
\end{figure}

\begin{figure}[h]
\vskip 3cm
\epsfig{figure=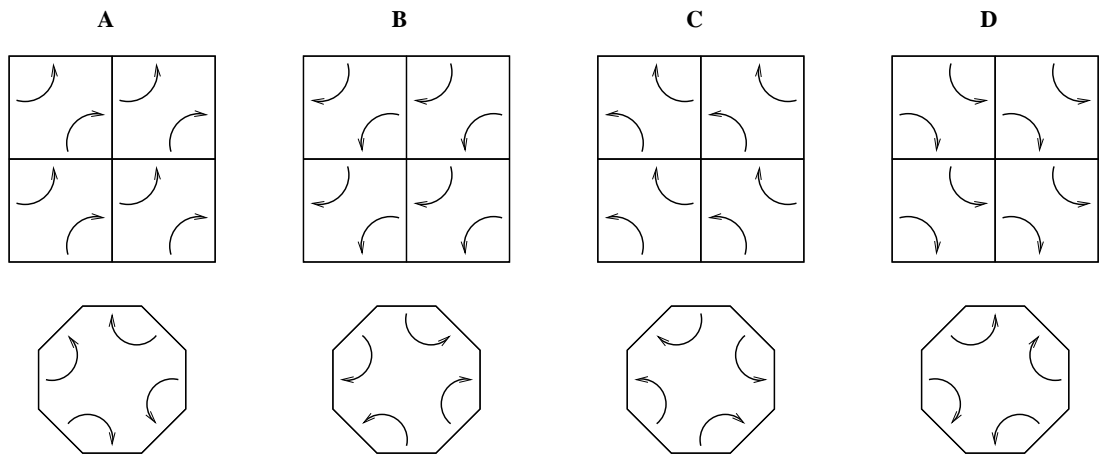,width=15cm}
\vskip 1cm
\caption{Graphical description of the action of the discrete Hodge operator on the edge values of lattice differentials. Four different definitions, related by $\pi /2$ rotations,  are possible. The definition given in the text (\ref{star}) corresponds to the drawing A. The four choices  have similar properties and generate the same period matrix.} 
\label{merd}
\end{figure}

\begin{figure}[h]
\vskip 3cm
\epsfig{figure=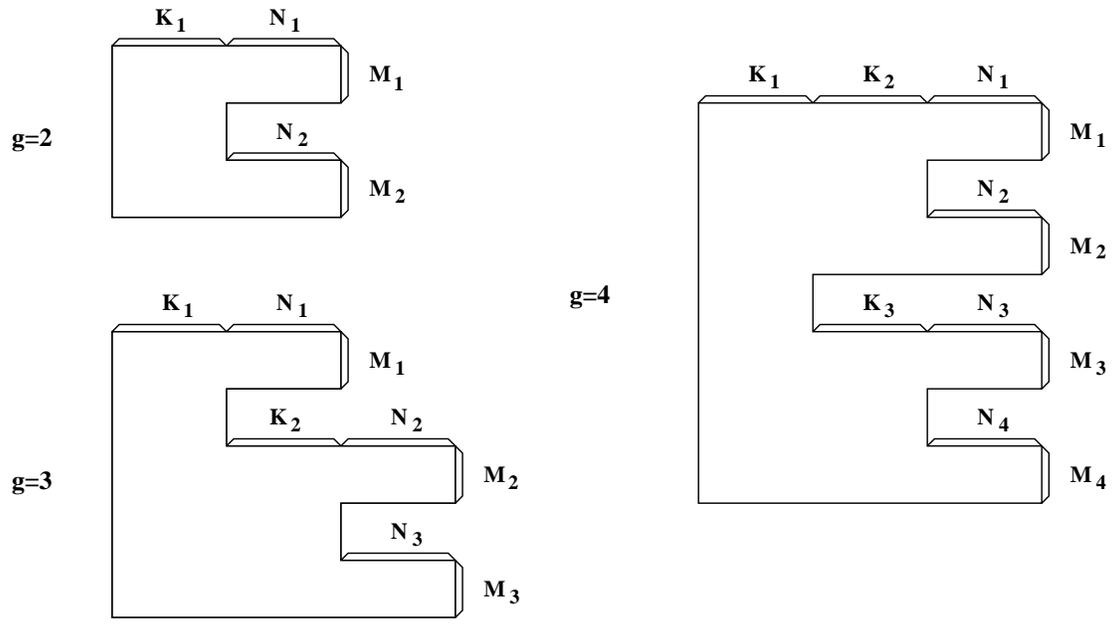,width=15cm}
\vskip 1cm
\caption{The integer sizes characterizing pair of pants decompositions of higher genus lattices for genus$=2,3$ and 4.}
\label{fig8}
\end{figure}

\begin{figure}[h]
\vskip 3cm
\epsfig{figure=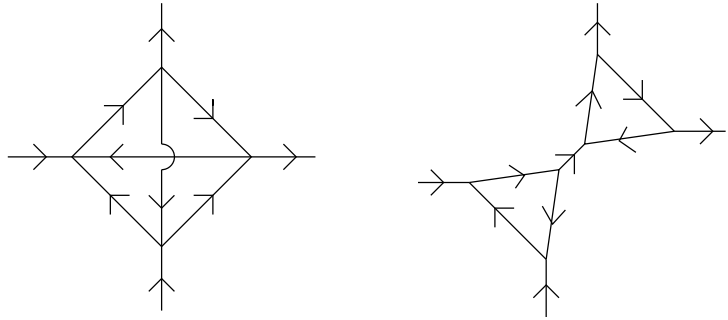,width=10cm}
\vskip 1cm
\caption{The Kasteleyn four vertices decoration and the Fisher six vertices decoration for the Ising model with four near neighbors.}
\label{fig99}
\end{figure}

\begin{figure}[h]
\vskip 3cm
\epsfig{figure=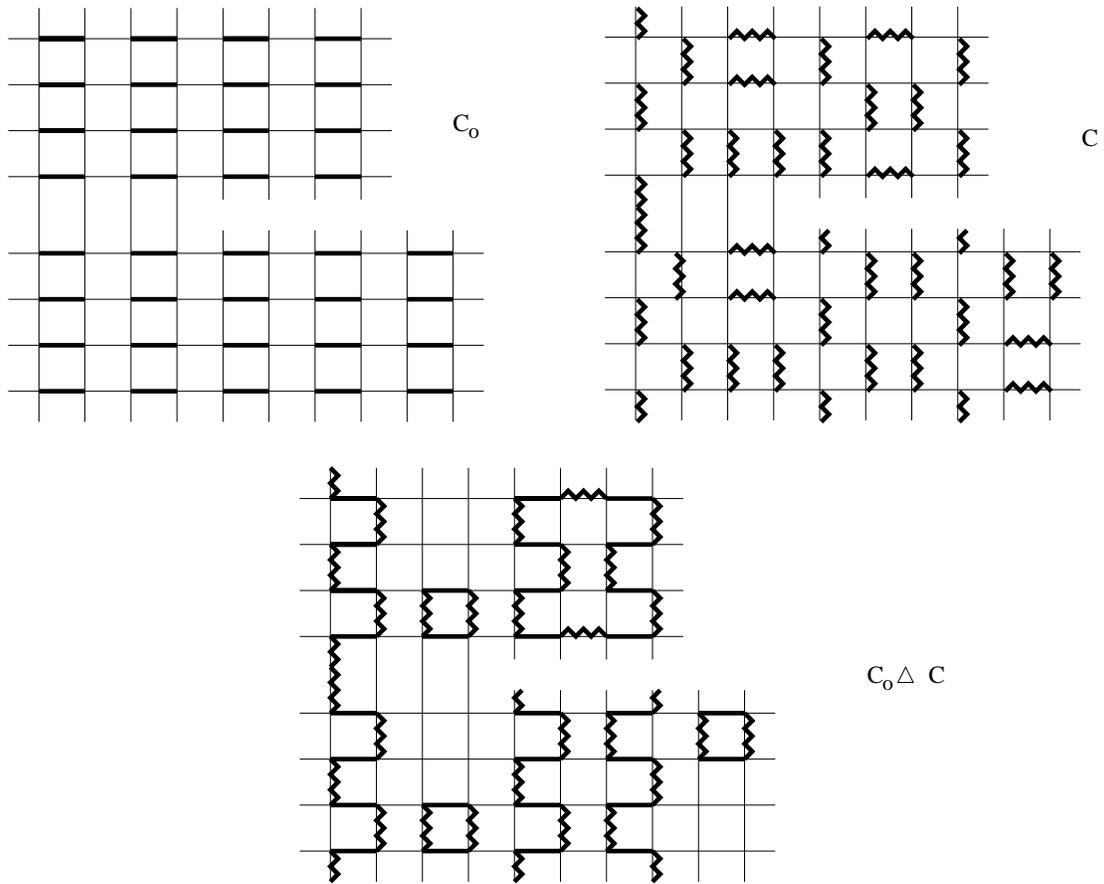,width=15cm}
\vskip 1cm
\caption{Two dimer configurations: the standard dimer configuration $C_0$ and an arbitrary dimer configuration $C$. The overlap of the two configurations with  double edges removed classifies $C$ to be of type $T(0102)=T(eoee)$.}
\label{fig9}
\end{figure}

\begin{figure}[h]
\vskip 3cm
\epsfig{figure=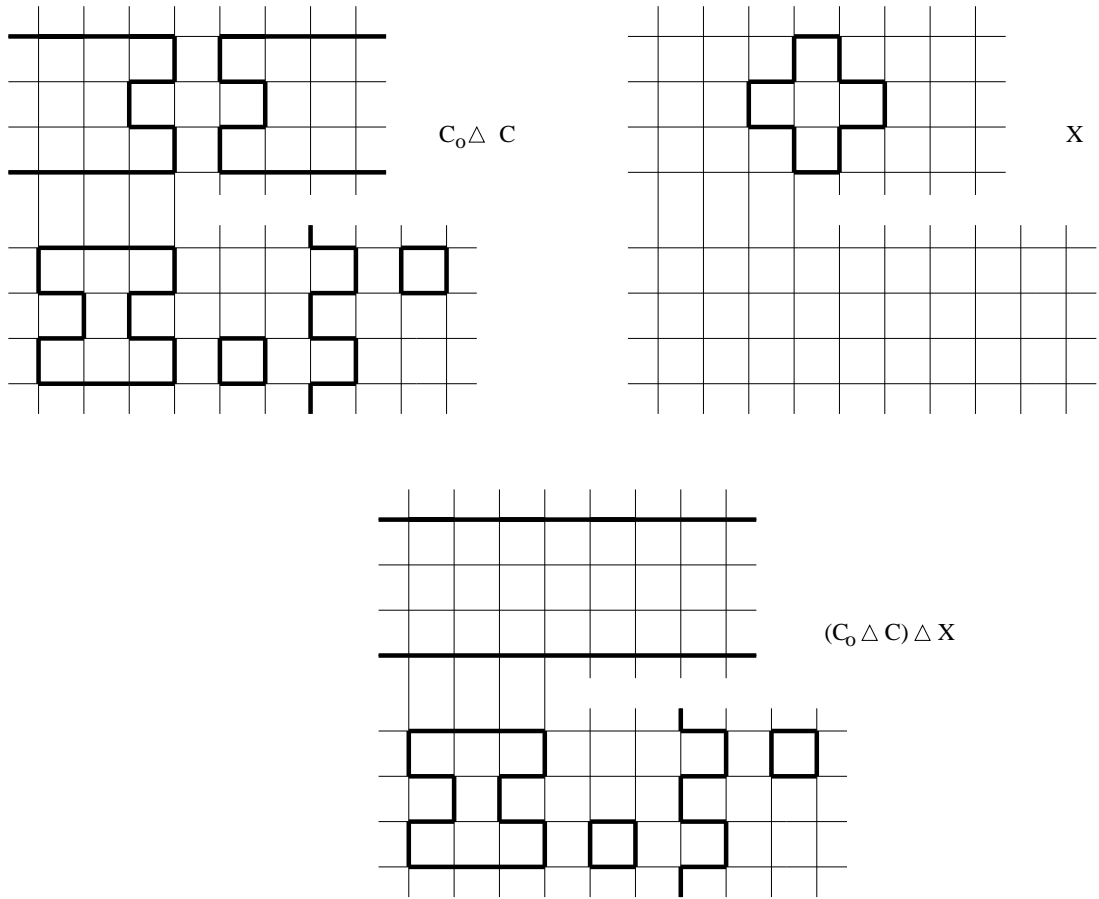,width=15cm}
\vskip 1cm
\caption{An overlap  of type $T(2001)$ can be built by overlapping a type $T(0001)$ with a boundary $X$ and therefore both belong to the same type $T(eeeo)$.}
\label{fig10}
\end{figure}

\begin{figure}[h]
\vskip 3cm
\epsfig{figure=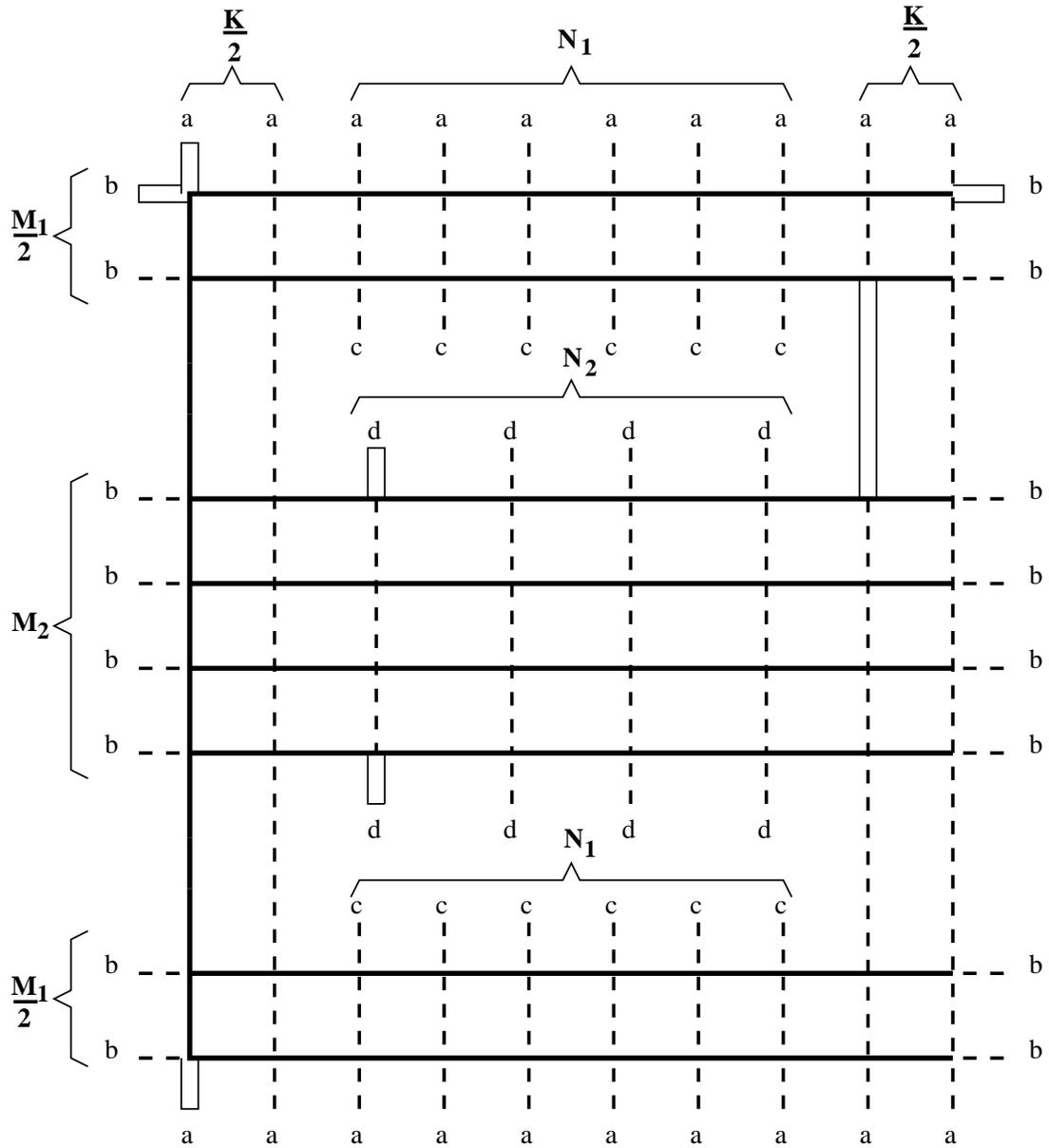,width=15cm}
\vskip 1cm
\caption{Graphical proof of the existence of 16 independent clockwise odd orientations of the genus two lattice. The lattice is redrawn with oriented edge identifications given by the letters a to d. The full line gives a spanning tree on the lattice edges were, using the local equivalence transformation, we can choose the orientation. On the dashed edges the orientation is determined by the clockwise odd condition and on the four rectangular edges the orientation is not determined. This gives a total of $2^4$ independent clockwise odd orientations. }
\label{fig11}
\end{figure}

Dimer configurations of type $T(eeee)$ will have the same sign on all clockwise odd orientations since its overlap with the standard configuration $C_0$ is topologically trivial.
The sign of a general configuration of type  $T(n_{b_1}n_{a_1}n_{b_2}n_{a_2})$ under the edge orientation $A(0000)$ can be evaluated by inspection on Fig. \ref{fig2} and \ref{fig4} after choosing a representative overlap diagram of each type, remembering that we are considering all the sizes $(M_1,M_2,K,N_1,N_2)$ to be  even integers. This is the last column of table \ref{table2}. 
Once these signs are know the remaining signs on the table \ref{table2} can be evaluated in a simple way: a relative sign of a configuration changes with the introduction of a disorder loop  if that configuration has a non-trivial transition cycle intersecting that disorder loop.

Once all the relative signs are known it is simple to find the  linear combination of Pfaffians that will give the same overall sign to all dimer configurations and we obtain equation (\ref{zzz}) and the last row of the matrix $b_{ij}$ in equation (\ref{zi}).  Equation (\ref{zzz})  can be rewritten using the notation of (\ref{defori}) as
\be
     Z(0000)= \frac{1}{4} \sum_{n_X=0,1}\; \alpha_{n_{\tilde b_1}n_{\tilde a_1}n_{\tilde b_2}n_{\tilde a_2}} \; \Pf(A(n_{\tilde b_1}n_{\tilde a_1}n_{\tilde b_2}n_{\tilde a_2}))
\ee
 with some choice of $\alpha_i=\pm 1$. 
The  Pfaffian expansion of the remaining partition functions $Z_i$ can be obtained form this equation  by successive inversions $0\rightarrow 1, 1\rightarrow 0$ on the same entry  $n_x$ on both the $Z()$ and the $A()$ terms and noticing that this correspond to a permutation of the $\alpha_i$. In this way we can construct the full matrix $b_{ij}$ that for our lattice, and labeling of the orientations, is given in table \ref{table3}.

\end{appendix}

\centering

\pagebreak

\pagebreak

\pagebreak

\pagebreak

\pagebreak

\pagebreak

\pagebreak

\pagebreak

\pagebreak

\pagebreak

\pagebreak

\pagebreak

\pagebreak

\pagebreak

%   ISING  convergence

\begin{table} \centering
\begin{tabular}{|c|ccccccc|}
L       & 2        &         3  &         4  &         5  &         6 &  7      & $\infty$\\\hline
$\cal N$    &     448  &      1008  &      1792  &      2800  &      4032 & 5488     &  extrapolated\\\hline
$R^2_1 $&0.001419  &  0.001746  &  0.001867  &  0.001923  &  0.001953 & 0.001972 & 0.002023\\
$R^2_2 $&0.000007  &  0.000005  &  0.000003  &  0.000002  &  0.000002 & 0.000001 & $10^{-7}$\\
$R^2_3 $&0.000000  &  0.000000  &  0.000000  &  0.000000  &  0.000000 & 0.000000 & $10^{-8}$\\
$R^2_4 $&0.000004  &  0.000003  &  0.000002  &  0.000001  &  0.000001 & 0.000001 & $10^{-7}$\\
$R^2_5 $&0.000003  &  0.000002  &  0.000001  &  0.000001  &  0.000001 & 0.000001 & $10^{-8}$\\
$R^2_6 $&0.004802  &  0.005766  &  0.006120  &  0.006284  &  0.006375 & 0.006432 & 0.006584\\
$R^2_7 $&0.000001  &  0.000001  &  0.000001  &  0.000001  &  0.000001 & 0.000001 & 0.000001\\
$R^2_8 $&0.003146  &  0.003899  &  0.004182  &  0.004316  &  0.004390 & 0.004437 & 0.004562\\
$R^2_9 $&0.000213  &  0.000117  &  0.000072  &  0.000048  &  0.000034 & 0.000026 & 0.000001\\
$R^2_{10}$&0.631982  &  0.657856  &  0.666786  &  0.670990  &  0.673343 & 0.674812 & 0.678782\\
$R^2_{11}$&0.311574  &  0.315561  &  0.316467  &  0.316709  &  0.316744 & 0.316704 & 0.316655\\
$R^2_{12}$&0.991780  &  0.992740  &  0.993053  &  0.993190  &  0.993260 & 0.993301 & 0.993414\\
$R^2_{13}$&0.000212  &  0.000116  &  0.000072  &  0.000048  &  0.000034 & 0.000026 & 0.000001\\
$R^2_{14}$&0.637340  &  0.662754  &  0.671528  &  0.675663  &  0.677981 & 0.679430 & 0.683345\\
$R^2_{15}$&0.309128  &  0.313357  &  0.314346  &  0.314625  &  0.314679 & 0.314650 & 0.314633\\
$R^2_{16}$&1.000000  &  1.000000  &  1.000000  &  1.000000  &  1.000000 & 1.000000 & 1.000000 \\
\end{tabular}
\caption{ Convergence with lattice size $\cal N$ of the ratios of determinants of the adjacency matrices  $R_i=P^2_i/P^2_{16}$ for the critical ISING model in a lattice with  $w_h$=0.66403677026784896368 $w_v $=0.20189651799465540849 and aspect ratio ($m_1$,$m_2$,k,$n_1$,$n_2$)=(4,2,2,2,4). See section \ref{section3} for discussion.}
\label{table4}
\end{table}

%   DIMERS convergence

\begin{table} \centering
\begin{tabular}{|c|ccccccccc|}
L    &         4  &         5  &         6  &         7  &         8  &         9  &        10  &        11 & $\infty$\\\hline
$\cal N$ &       640  &      1000  &      1440  &      1960  &      2560  &      3240  &      4000  &      4840 &  extrapolated\\ \hline
$R_1$&  0.209606  &  0.213826  &  0.215781  &  0.216790  &  0.217348  &  0.217669  &  0.217855  &  0.217961 & 0.218541   \\
$R_2$&  0.000024  &  0.000017  &  0.000012  &  0.000009  &  0.000007  &  0.000006  &  0.000005  &  0.000004 & $10^{-7}$ \\
$R_3$&  0.000340  &  0.000243  &  0.000181  &  0.000139  &  0.000110  &  0.000089  &  0.000073  &  0.000061 & 0.000004   \\
$R_4$&  0.000010  &  0.000007  &  0.000005  &  0.000004  &  0.000003  &  0.000003  &  0.000002  &  0.000002 & $10^{-7}$\\
$R_5$&  0.000017  &  0.000012  &  0.000008  &  0.000006  &  0.000005  &  0.000004  &  0.000003  &  0.000003 & $10^{-7}$ \\
$R_6$&  0.805524  &  0.808824  &  0.810645  &  0.811765  &  0.812510  &  0.813033  &  0.813417  &  0.813709 & 0.815064   \\
$R_7$&  0.000024  &  0.000024  &  0.000024  &  0.000023  &  0.000023  &  0.000023  &  0.000023  &  0.000023  & 0.000022  \\
$R_8$&  0.555230  &  0.569831  &  0.577729  &  0.582534  &  0.585711  &  0.587939  &  0.589574  &  0.590815 & 0.596530   \\
$R_9$&  0.000457  &  0.000325  &  0.000240  &  0.000184  &  0.000145  &  0.000117  &  0.000097  &  0.000081  & 0.000004  \\
$R_{10}$&  0.123728  &  0.127266  &  0.129148  &  0.130271  &  0.130998  &  0.131498  &  0.131857  &  0.132125 & 0.133375  \\
$R_{11}$&  0.262676  &  0.266688  &  0.268415  &  0.269229  &  0.269627  &  0.269816  &  0.269893  &  0.269907 & 0.270077 \\
$R_{12}$&  0.183361  &  0.184107  &  0.184460  &  0.184642  &  0.184739  &  0.184791  &  0.184818  &  0.184830 & 0.184916   \\
$R_{13}$&  0.000116  &  0.000082  &  0.000060  &  0.000046  &  0.000036  &  0.000029  &  0.000024  &  0.000020 & 0.000001 \\
$R_{14}$&  0.686459  &  0.701963  &  0.710275  &  0.715308  &  0.718626  &  0.720950  &  0.722654  &  0.723949 & 0.729883 \\
$R_{15}$&  0.050147  &  0.050986  &  0.051336  &  0.051492  &  0.051559  &  0.051584  &  0.051586  &  0.051576 & 0.051561 \\
$R_{16}$&  1.000000  &  1.000000  &  1.000000  &  1.000000  &  1.000000  &  1.000000  &  1.000000  &  1.000000 & 1.000000 \\
\end{tabular}
\caption{Convergence with lattice size $\cal N$ of the ratios of determinants of the adjacency matrices for the DIMER model in a lattice with  $z_h$= 0.780 $z_v$=0.560 and aspect ratio ($m_1$,$m_2$,k,$n_1$,$n_2$)=(4,2,2,6,2). Additional discussion is given in section \ref{section3}.}
\label{table5}
\end{table}

%   ISING versus theta

\begin{table} \centering
\begin{tabular}{|c|ccc|ccc|ccc|}
 & \multicolumn{3}{c|}{\textbf{I1}: [0.537,0.301] (2,2,2,2,2)}   & \multicolumn{3}{c|}{\textbf{I2}: [0.335,0.498](4,4,2,2,2)}   & \multicolumn{3}{c|}{\textbf{I3}: [0.664,0.202](4,2,2,2,4)}  \\ \hline
i &   $R^2_i$ &$\theta^4_i$(fit)&$\theta^4_i$(eval)  &   $R^2_i$ &$\theta^4_i$(fit)&$\theta^4_i$(eval) &   $R^2_i$ &$\theta^4_i$(fit)&$\theta^4_i$(eval)\\ \hline
   1&  0.214707&  0.214705&  0.214872&  0.214748&  0.214751&  0.214917&  0.002023&  0.002023&  0.002015 \\
   2&  0.000001&  0.000000&  0.000000&  0.000002&  0.000000&  0.000000&  0.000000&  0.000000&  0.000000 \\
   3&  0.000001&  0.000000&  0.000000&  0.000001&  0.000000&  0.000000&  0.000000&  0.000000&  0.000000 \\
   4&  0.000001&  0.000000&  0.000000&  0.000001&  0.000000&  0.000000&  0.000000&  0.000000&  0.000000 \\
   5&  0.000000&  0.000000&  0.000000&  0.000001&  0.000000&  0.000000&  0.000000&  0.000000&  0.000000 \\
   6&  0.664619&  0.664616&  0.663636&  0.664364&  0.664366&  0.663404&  0.006584&  0.006585&  0.006544 \\
   7&  0.000339&  0.000338&  0.000350&  0.000341&  0.000341&  0.000353&  0.000001&  0.000001&  0.000001 \\
   8&  0.450251&  0.450250&  0.449115&  0.449951&  0.449957&  0.448840&  0.004562&  0.004563&  0.004529 \\
   9&  0.000001&  0.000000&  0.000000&  0.000002&  0.000000&  0.000000&  0.000001&  0.000000&  0.000000 \\
  10&  0.214707&  0.214705&  0.214880&  0.214748&  0.214751&  0.214917&  0.678782&  0.678782&  0.677686 \\
  11&  0.335047&  0.335045&  0.336005&  0.335287&  0.335292&  0.336243&  0.316655&  0.316655&  0.317785 \\
  12&  0.335047&  0.335045&  0.336014&  0.335287&  0.335292&  0.336243&  0.993414&  0.993414&  0.993456 \\
  13&  0.000000&  0.000000&  0.000000&  0.000001&  0.000000&  0.000000&  0.000001&  0.000000&  0.000000 \\
  14&  0.664619&  0.664616&  0.663645&  0.664364&  0.664366&  0.663404&  0.683345&  0.683344&  0.682214 \\
  15&  0.120683&  0.120679&  0.121484&  0.120883&  0.120883&  0.121679&  0.314633&  0.314633&  0.315771 \\
  16&  1.000000&  1.000000&  1.000000&  1.000000&  1.000000&  1.000000&  1.000000&  1.000000&  1.000000 \\\hline\hline
 &$\Omega_{11}$&$\Omega_{12}$&$\Omega_{22}$&$\Omega_{11}$&$\Omega_{12}$&$\Omega_{22}$&$\Omega_{11}$&$\Omega_{12}$&$\Omega_{22}$ \\\hline 
 $\Omega$(fit)&   1.174907&   $-$1.024878&  2.049756&   1.174646&   $-$1.024365&  2.048730&   0.403512&   $-$0.368642&   1.526888  \\ 
 $\Omega$(eval)&  1.17389&   $-$1.02281&  2.04563&   1.17365&   $-$1.02232&  2.04464&   0.40320&   $-$0.367695&   1.52404\\
\end{tabular}
\caption{Comparison of ratios of determinants of adjacency matrices with ratios of theta functions for the critical ISING model in three different lattices: I1, I2 and I3.  Each lattice is characterized by the couplings and aspect ratio [$w_h $,$w_v $]($m_1$,$m_2$,k,$n_1$,$n_2$). The $L\rightarrow \infty$ ratios are compared with theta function ratios for a fitted period matrix $\Omega$(fit) and the period matrix  $\Omega$(eval) evaluated by the procedure of section \ref{section4} (modular choice A). The first two sets of values illustrate the fact that re-scaling of the couplings can be compensated by a re-scaling of the overall vertical/horizontal aspect ratio.} 
%A   .53704956699803528586  .30119421191220209666  .537 .301 (2,2,2,2,2)
%B   .33501238283186393250  .49811344502853716476  .335 .498 (4,4,2,2,2)
%C   .66403677026784896368  .20189651799465540849  .664 .202 (4,2,2,2,4)
\label{table8}
\end{table}

%   DIMER versus theta

\begin{table} \centering
\begin{tabular}{|c|ccc|ccc|ccc|} 
 & \multicolumn{3}{c|}{\textbf{D1}: [1.100,0.900](2,2,2,2,2)} &\multicolumn{3}{c|}{\textbf{D2}: [0.275,0.450](4,4,2,2,2)} &\multicolumn{3}{c|}{\textbf{D3}: [0.780,0.560](4,2,2,6,2)}\\ \hline 
i &   $R_i$ &$\theta^4_i$(fit)&$\theta^4_i$(eval)& $R_i$ &$\theta^4_i$(fit)& $\theta^4_i$(eval)&  $R_i$ &$\theta^4_i$(fit)& $\theta^4_i$(eval)\\ \hline
  1 &   0.136210 &   0.136207 &   0.136552 &   0.136168 &   0.136061 &   0.136671 &   0.218541 &   0.218546 &   0.219812  \\
  2 &   0.000001 &   0.000000 &   0.000000 &   0.000001 &   0.000000 &   0.000000 &   0.000000 &   0.000000 &   0.000000  \\
  3 &   0.000001 &   0.000000 &   0.000000 &   0.000001 &   0.000000 &   0.000000 &   0.000004 &   0.000000 &   0.000000  \\
  4 &   0.000001 &   0.000000 &   0.000000 &   0.000001 &   0.000000 &   0.000000 &   0.000000 &   0.000000 &   0.000000  \\
  5 &   0.000000 &   0.000000 &   0.000000 &   0.000000 &   0.000000 &   0.000000 &   0.000000 &   0.000000 &   0.000000  \\
  6 &   0.828760 &   0.828757 &   0.828055 &   0.828291 &   0.828547 &   0.827822 &   0.815064 &   0.815061 &   0.814718  \\
  7 &   0.000385 &   0.000384 &   0.000397 &   0.000387 &   0.000405 &   0.000402 &   0.000022 &   0.000022 &   0.000023  \\
  8 &   0.692938 &   0.692934 &   0.691901 &   0.693182 &   0.692891 &   0.691553 &   0.596530 &   0.596536 &   0.594928  \\
  9 &   0.000001 &   0.000000 &   0.000000 &   0.000001 &   0.000000 &   0.000000 &   0.000004 &   0.000000 &   0.000000  \\
 10 &   0.136210 &   0.136207 &   0.136552 &    0.13628 &   0.136213 &   0.136666 &   0.133375 &   0.133377 &   0.133252  \\
 11 &   0.170862 &   0.170859 &   0.171547 &   0.170903 &   0.170896 &   0.171781 &   0.270077 &   0.270087 &   0.271820  \\
 12 &   0.170862 &   0.170859 &   0.171547 &   0.171024 &   0.171048 &   0.171776 &   0.184916 &   0.184917 &   0.185259  \\
 13 &   0.000000 &   0.000000 &   0.000000 &   0.000000 &   0.000000 &   0.000000 &   0.000001 &   0.000000 &   0.000000  \\
 14 &   0.828760 &   0.828757 &   0.828055 &   0.828514 &   0.828699 &   0.827818 &   0.729883 &   0.729892 &   0.728157  \\
 15 &   0.035039 &   0.035036 &   0.035393 &   0.035084 &    0.03524 &   0.035512 &   0.051561 &   0.051562 &   0.052031  \\
 16 &   1.000000 &   1.000000 &   1.000000 &   1.000000 &   1.000000 &   1.000000 &   1.000000 &   1.000000 &   1.000000  \\\hline\hline
 &$\Omega_{11}$&$\Omega_{12}$&$\Omega_{22}$&$\Omega_{11}$&$\Omega_{12}$&$\Omega_{22}$&$\Omega_{11}$&$\Omega_{12}$&$\Omega_{22}$ \\\hline 
 $\Omega$(fit)&  1.422286&   $-$1.208265&   2.416530& 1.422067&   $-$1.205355&   2.411027& 1.389893&   $-$1.297942&   2.457896 \\ 
 $\Omega$(eval)&  1.42099&   $-$1.20562&   2.41124&  1.42056&   $-$1.20477&   2.40953&  1.38931&   $-$1.29614&   2.45252   \\ 
\end{tabular}
\caption{ Comparison of  ratios of determinants of adjacency matrices  $R_i$ with ratios of theta functions for the DIMER model in three different lattices: P1, P2 and P3. Each lattice is characterized by the dimer weights and aspect ratio [$z_h$,$z_v$]($m_1$,$m_2$,k,$n_1$,$n_2$).} 
%A  1.100  0.900  (2,2,2,2,2)
%B  0.275  0.450  (4,4,2,2,2)
%C  0.780  0.560  (4,2,2,6,2)
\label{table9}
\end{table}

%  period matrix convergence and modular properties... more digits

\begin{table}\centering 
\begin{tabular}{|c|c|c|c|c|c|c|c|c|c|c|}
\multicolumn{2}{|c|}{}& \multicolumn{3}{c|}{\textbf{A}: (4,2,2,6,2)} &\multicolumn{3}{c|}{\textbf{B}: (4,2,2,6,2) with $a_i \leftrightarrow b_i$}&\multicolumn{3}{c|}{\textbf{C}: (2,4,2,2,6)} \\ \hline
   L  & $\cal N$  &$\Omega_{11}$&$\Omega_{12}$&$\Omega_{22}$&$\Omega_{11}$&$\Omega_{12}$&$\Omega_{22}$&$\Omega_{11}$&$\Omega_{12}$&$\Omega_{22}$ \\\hline 
   5 &  1000&   1.383663 &  $-$1.279193 &   2.401681 &   1.414286 &   0.7392862 &    0.7821439 &   1.226958 &  $-$1.122488 &   2.401681 \\
   6 &  1440&   1.384849 &  $-$1.282726 &   2.412281 &   1.415022 &   0.7407580 &    0.7850874 &   1.231669 &  $-$1.129555 &   2.412281 \\
   7 &  1960&   1.385681 &  $-$1.285248 &   2.419848 &   1.415555 &   0.7418236 &    0.7872187 &   1.235032 &  $-$1.134599 &   2.419848 \\
   8 &  2560&   1.386311 &  $-$1.287139 &   2.425518 &   1.415958 &   0.7426305 &    0.7888325 &   1.237552 &  $-$1.138380 &  2.425518 \\
   9 &  3240&   1.386801 &  $-$1.288608 &   2.429926 &   1.416274 &   0.7432626 &    0.7900967 &   1.239511 &  $-$1.141318 &  2.429926 \\
  10 &  4000&   1.387192 &  $-$1.289782 &   2.433450 &   1.416528 &   0.7437711 &    0.7911136 &   1.241077 &  $-$1.143667 &  2.433450 \\
  11&   4840&   1.387513 &  $-$1.290743 &   2.436332 &   1.416737 &   0 7441889 &    0.7919492 &   1.242358 &  $-$1.145589 &  2.436332 \\ \hline
 $\infty$ &  $\infty$ &  1.38931  &  $-1$.29614  &  2.45252   &  1.41792  & 0.746553  & 0.796677  &  1.24955  & $-$1.15638  & 2.45252 \\
\end{tabular}
\caption{ Convergence  with lattice size $\cal N$ of the period matrix elements for a  lattice with h=0.780 v=0.560 and aspect ratio ($m_1$,$m_2$,k,$n_1$,$n_2$)=(4,2,2,6,2) for  three different choices of the homology basis: A, B and C. See section \ref{section6} for discussion.}
\label{table6}
\end{table}

\begin{table} \centering
\begin{tabular}{|c|c|c|c|c|c|c|c|c|}
 \multicolumn{3}{|c|}{\textbf{A}: (4,2,2,6,2)} &\multicolumn{3}{c|}{\textbf{B}: (4,2,2,6,2) with $a_i \leftrightarrow b_i$}&\multicolumn{3}{c|}{\textbf{C}: (2,4,2,2,6)} \\ \hline
    i    &   $\theta_i(4840)$ &   $\theta_i(\infty)$  &   i    &   $\theta_i(4840)$ &   $\theta_i(\infty)$  &    i    &   $\theta_i(4840)$ &   $\theta_i(\infty)$  \\\hline
  1 & 0.223711 & 0.219812 &  1 & 0.210138 & 0.213846 & 10 & 0.223711 & 0.219812 \\
  2 & 0.000000 & 0.000000 &  3 & 0.000000 & 0.000000 &  9 & 0.000000 & 0.000000 \\
  3 & 0.000000 & 0.000000 &  2 & 0.000000 & 0.000000 &  4 & 0.000000 & 0.000000 \\
  4 & 0.000000 & 0.000000 &  4 & 0.000000 & 0.000000 &  3 & 0.000000 & 0.000000 \\
  5 & 0.000000 & 0.000000 &  9 & 0.000000 & 0.000000 & 13 & 0.000000 & 0.000000  \\
  6 & 0.813656 & 0.814718 & 11 & 0.817352 & 0.816342 & 14 & 0.813656 & 0.814718 \\
  7 & 0.000027 & 0.000023 & 10 & 0.000014 & 0.000017 &  7 & 0.000027 & 0.000023 \\
  8 & 0.589972 & 0.594928 & 12 & 0.607228 & 0.602514 &  8 & 0.589972 & 0.594928 \\
  9 & 0.000000 & 0.000000 &  5 & 0.000000 & 0.000000 &  2 & 0.000000 & 0.000000 \\
 10 & 0.132858 & 0.133252 &  7 & 0.134120 & 0.133805 &  1 & 0.132858 & 0.133252 \\
 11 & 0.277170 & 0.271820 &  6 & 0.258651 & 0.263681 & 12 & 0.277170 & 0.271820 \\
 12 & 0.186317 & 0.185259 &  8 & 0.182634 & 0.183640 & 11 & 0.186317 & 0.185259 \\
 13 & 0.000000 & 0.000000 & 13 & 0.000000 & 0.000000 &  5 & 0.000000 & 0.000000 \\
 14 & 0.722803 & 0.728157 & 15 & 0.741334 & 0.736302 &  6 & 0.722803 & 0.728157 \\
 15 & 0.053486 & 0.052031 & 14 & 0.048528 & 0.049852 & 15 & 0.053486 & 0.052031 \\
 16 & 1.000000 & 1.000000 & 16 & 1.000000 & 1.000000 & 16 & 1.000000 & 1.000000 \\
\end{tabular}
\caption{The sixteen theta function ratios for the three period matrices calculations of table \ref{table6}, both the largest size and the extrapolated value are given. The different ratios are rearranged to facilitate comparison.}
\label{table7}
\end{table}

\begin{table}\centering 
\begin{tabular}{|c|cccccccccccccccc|}
 &$A_1$ &$A_2$ &$A_3$ &$A_4$ &$A_5$ &$A_6$ &$A_7$ &$A_8$ &$A_9$&$A_{10}$&$A_{11}$&$A_{12}$&$A_{13}$&$A_{14}$&$A_{15}$&$A_{16}$ \\ \hline
%      &(++++)&(+++-)&(++-+)&(++--)&(+-++)&(+-+-)&(+--+)&(+---)&(-+++)&(-++-)&(-+-+)&(-+--)&(--++)&(--+-)&(---+)&(----)\\
%  & (----)   & (---+)   & (--+-)   & (--++)   & (-+--)   & (-+-+)   & (-++-)   & (-+++)   & (+---)   & (+--+)   & (+-+-)   & (+-++)   & (++--)   & (++-+)   & (+++-)   & (++++)  \\
(eeee)	 & +	 & +	 & +	 & +	 & +	 & +	 & +	 & +	 & +	 & +	 & +	 & +	 & +	 & +	 & +	 & +	\\
(eeeo)	 & $-$	 & +	 & $-$	 & +	 & $-$	 & +	 & $-$	 & +	 & $-$	 & +	 & $-$	 & +	 & $-$	 & +	 & $-$	 & +	\\
(eeoe)	 & $-$	 & $-$	 & +	 & +	 & $-$	 & $-$	 & +	 & +	 & $-$	 & $-$	 & +	 & +	 & $-$	 & $-$	 & +	 & +	\\
(eeoo)	 & $-$	 & +	 & +	 & $-$	 & $-$	 & +	 & +	 & $-$	 & $-$	 & +	 & +	 & $-$	 & $-$	 & +	 & +	 & $-$	\\
(eoee)	 & $-$	 & $-$	 & $-$	 & $-$	 & +	 & +	 & +	 & +	 & $-$	 & $-$	 & $-$	 & $-$	 & +	 & +	 & +	 & +	\\
(eoeo)	 & +	 & $-$	 & +	 & $-$	 & $-$	 & +	 & $-$	 & +	 & +	 & $-$	 & +	 & $-$	 & $-$	 & +	 & $-$	 & +	\\
(eooe)	 & +	 & +	 & $-$	 & $-$	 & $-$	 & $-$	 & +	 & +	 & +	 & +	 & $-$	 & $-$	 & $-$	 & $-$	 & +	 & +	\\
(eooo)	 & +	 & $-$	 & $-$	 & +	 & $-$	 & +	 & +	 & $-$	 & +	 & $-$	 & $-$	 & +	 & $-$	 & +	 & +	 & $-$	\\
(oeee)	 & $-$	 & $-$	 & $-$	 & $-$	 & $-$	 & $-$	 & $-$	 & $-$	 & +	 & +	 & +	 & +	 & +	 & +	 & +	 & +	\\
(oeeo)	 & +	 & $-$	 & +	 & $-$	 & +	 & $-$	 & +	 & $-$	 & $-$	 & +	 & $-$	 & +	 & $-$	 & +	 & $-$	 & +	\\
(oeoe)	 & +	 & +	 & $-$	 & $-$	 & +	 & +	 & $-$	 & $-$	 & $-$	 & $-$	 & +	 & +	 & $-$	 & $-$	 & +	 & +	\\
(oeoo)	 & +	 & $-$	 & $-$	 & +	 & +	 & $-$	 & $-$	 & +	 & $-$	 & +	 & +	 & $-$	 & $-$	 & +	 & +	 & $-$	\\
(ooee)	 & $-$	 & $-$	 & $-$	 & $-$	 & +	 & +	 & +	 & +	 & +	 & +	 & +	 & +	 & $-$	 & $-$	 & $-$	 & $-$	\\
(ooeo)	 & +	 & $-$	 & +	 & $-$	 & $-$	 & +	 & $-$	 & +	 & $-$	 & +	 & $-$	 & +	 & +	 & $-$	 & +	 & $-$	\\
(oooe)	 & +	 & +	 & $-$	 & $-$	 & $-$	 & $-$	 & +	 & +	 & $-$	 & $-$	 & +	 & +	 & +	 & +	 & $-$	 & $-$	\\
(oooo)	 & +	 & $-$	 & $-$	 & +	 & $-$	 & +	 & +	 & $-$	 & $-$	 & +	 & +	 & $-$	 & +	 & $-$	 & $-$	 & +	\\ \hline
sign	 & +	 & $-$	 & $-$	 & $-$	 & $-$	 & +	 & +	 & +	 & $-$	 & +	 & +	 & +	 & $-$	 & +	 & +	 & +	\\
\end{tabular}
\caption{ The relative signs for each dimer configuration type in the various clockwise odd edge orientations. The linear combination of the orientations that gives the same sign to all configurations is shown in the last row.}
\label{table2}
\end{table}

\begin{table} \centering
\begin{tabular}{|r|rrrrrrrrrrrrrrrr|}
  &$P_1$ &$P_2$ &$P_3$ &$P_4$ &$P_5$ &$P_6$ &$P_7$ &$P_8$ &$P_9$&$P_{10}$&$P_{11}$&$P_{12}$&$P_{13}$&$P_{14}$&$P_{15}$&$P_{16}$ \\ \hline
%abab  &(++++)&(+++-)&(++-+)&(++--)&(+-++)&(+-+-)&(+--+)&(+---)&(-+++)&(-++-)&(-+-+)&(-+--)&(--++)&(--+-)&(---+)&(----)&      \\
%1122
$Z_{1}$&1&1&1& $-1$ &1&1&1& $-1$&1&1&1& $-1$& $-1$& $-1$& $-1$&1\\
$Z_{2}$&1&1& $-1$&1&1&1& $-1$&1&1&1& $-1$&1& $-1$& $-1$&1& $-1$\\
$Z_{3}$&1& $-1$&1&1&1& $-1$&1&1&1& $-1$&1&1& $-1$&1& $-1$& $-1$\\
$Z_{4}$& $-1$&1&1&1& $-1$&1&1&1& $-1$&1&1&1&1& $-1$& $-1$& $-1$\\
$Z_{5}$&1&1&1& $-1$ &1&1&1& $-1$& $-1$& $-1$& $-1$&1&1&1&1& $-1$\\
$Z_{6}$&1&1& $-1$&1&1&1& $-1$&1& $-1$& $-1$&1& $-1$&1&1& $-1$&1\\
$Z_{7}$&1& $-1$&1&1&1& $-1$&1&1& $-1$&1& $-1$& $-1$&1& $-1$&1&1\\
$Z_{8}$& $-1$&1&1&1& $-1$&1&1&1&1& $-1$& $-1$& $-1$& $-1$&1&1&1\\
$Z_{9}$&1&1&1& $-1$ & $-1$& $-1$& $-1$&1&1&1&1& $-1$&1&1&1& $-1$\\
$Z_{10}$&1&1& $-1$&1& $-1$& $-1$&1& $-1$&1&1& $-1$&1&1&1& $-1$&1\\
$Z_{11}$&1& $-1$&1&1& $-1$&1& $-1$& $-1$&1& $-1$&1&1&1& $-1$&1&1\\
$Z_{12}$& $-1$&1&1&1&1& $-1$& $-1$& $-1$& $-1$&1&1&1& $-1$&1&1&1\\
$Z_{13}$& $-1$& $-1$& $-1$&1&1&1&1& $-1$&1&1&1& $-1$&1&1&1& $-1$\\
$Z_{14}$& $-1$& $-1$&1& $-1$ &1&1& $-1$&1&1&1& $-1$&1&1&1& $-1$&1\\
$Z_{15}$& $-1$&1& $-1$& $-1$ &1& $-1$&1&1&1& $-1$&1&1&1& $-1$&1&1\\
$Z_{16}$&1& $-1$& $-1$& $-1$ & $-1$&1&1&1& $-1$&1&1&1& $-1$&1&1&1\\
\end{tabular}
\caption{ The b matrix relating the sixteen partition functions with the Pfaffians of the sixteen clockwise odd orientations. Each $Z_i$ is given by $\frac{1}{4}$ of the linear combination of the $P_j$ in the corresponding row.}
\label{table3}
\end{table}

\end{document}